\PassOptionsToPackage{hyphens}{url}
\PassOptionsToPackage{obeyspaces}{url}
\PassOptionsToPackage{spaces}{url}
\PassOptionsToPackage{dvipsnames}{xcolor}

\documentclass[conference]{IEEEtran}

\newcommand{\updated}[1]{#1}

\usepackage[english]{babel}
\usepackage[nolist,nohyperlinks]{acronym}
\usepackage{multirow}
\usepackage{wasysym}
\usepackage{subcaption}

\usepackage{url}

\newcommand{\del}[1]{}

\newcommand{\ignore}[1]{}
\newcommand{\ntg}[1]{}

\usepackage{xparse}
\usepackage{xstring}
\NewDocumentCommand{\rfc}{m o}{%
    \StrCount{#1}{,}[\commacount]
    \ifnum\commacount > 0%
        \def\rfcprefix{RFCs}
    \else%
        \def\rfcprefix{RFC}
    \fi%
    \IfValueTF{#2}{%
        \StrCount{#2}{,}[\commacount]
        \ifnum\commacount < 1%
            \def\secprefix{\S}
        \else%
            \def\secprefix{\S\S}
        \fi%
        \rfcprefix~#1~\secprefix#2%
    }{%
        \rfcprefix~#1%
    }%
}

\usepackage{array}
\makeatletter
\newcolumntype{"}{@{\hskip\tabcolsep\vrule width 1pt\hskip\tabcolsep}}
\newcolumntype{;}{@{\hskip\tabcolsep\vrule width 1pt}}
\makeatother
\usepackage{graphicx}

\usepackage{listings}
\usepackage{xcolor}
\usepackage{hyperref}
\lstdefinelanguage{ASN1}{
  morekeywords={SEQUENCE,INTEGER,DEFAULT,OF,OBJECT,IDENTIFIER,GeneralizedTime},
  sensitive=true,
}

\lstdefinelanguage{Protobuf}{
  morekeywords={
    syntax, message, repeated, optional,
    string, bytes, uint32, uint64, 
  },
  sensitive=true,
  morecomment=[l]{//},
  morestring=[b]{"},
}

\begin{document}

\begin{acronym}
    \acro{rpki}[RPKI]{Resource Public Key Infrastructure}
    \acroindefinite{rpki}{an}{a}

    \acro{bgp}[BGP]{Border Gateway Protocol}
    
    \acro{rrdp}[RRDP]{\acs{rpki} Repository Delta Protocol}
    \acroindefinite{rrdp}{an}{a}
    
    \acro{rtr}[RTR]{\acs{rpki} to Router Protocol}
    \acroindefinite{rtr}{an}{a}
    
    \acro{rpsl}[RPSL]{Routing Policy Specification Language}
    \acroindefinite{rpsl}{an}{a}
    
    \acro{rp}[RP]{Relying Party}
    \acroplural{rp}[RPs]{Relying Parties}
    \acroindefinite{rp}{an}{a}
    
    \acro{roa}[ROA]{Route Origin Authorization}
    \acroindefinite{roa}{an}{a}
    
    \acro{cms}[CMS]{Cryptographic Message Syntax}
    
    \acro{crl}[CRL]{Certificate Revocation List}
    
    \acro{ca}[CA]{Certificate Authority}
    \acroplural{ca}[CAs]{Certificate Authorities}
    
    \acro{as}[AS]{Autonomous System}
    \acroindefinite{as}{an}{an}
    
    \acro{pp}[PP]{Publication Point}
    
    \acro{rir}[RIR]{Regional Internet Registry}
    \acroplural{rir}[RIRs]{Regional Internet Registries}
    \acroindefinite{rir}{an}{a}
    
    \acro{lir}[LIR]{Local Internet Registry}
    \acroplural{lir}[LIRs]{Local Internet Registries}
    
    \acro{slurm}[SLURM]{Simplified Local Internet Number Resource Management}
    
    \acro{tal}[TAL]{Trust Anchor Locator}
    
    \acro{cp}[CP]{Certificate Policy}
    \acroplural{cp}[CPs]{Certificate Policies}
    
    \acro{rov}[ROV]{Route Origin Validation}
    \acroindefinite{rov}{an}{a}
    
    \acro{vrp}[VRP]{Validated \acs{roa} Payload}
    
    \acro{dos}[DoS]{denial-of-service}
\end{acronym}

\title{Pruning the Tree: Rethinking RPKI\\Architecture From The Ground Up}

\author{Haya Schulmann$^{\ddagger\S}$\qquad \qquad Niklas Vogel$^{\ddagger\S}$\vspace{2mm}\\
{\small
$^{\ddagger}$Goethe-Universität Frankfurt \qquad $^{\S}$ATHENE German Research Center for Applied Cybersecurity} \\
}

\maketitle

\acresetall 

\begin{abstract}

Resource Public Key Infrastructure (RPKI) is a critical security mechanism for BGP, but the complexity of its architecture is a growing concern as its adoption scales. Current RPKI design heavily reuses legacy PKI components, such as X.509 EE-certificates, ASN.1 encoding, and XML-based repository protocols, which introduce excessive cryptographic validation, redundant metadata, and inefficiencies in both storage and processing. We show that these design choices, although based on established standards, create significant performance bottlenecks, increase the vulnerability surface, and hinder scalability for wide-scale Internet deployment.

In this paper, we perform the first systematic analysis of the root causes of complexity in RPKI's design and experimentally quantify their real-world impact. We show that over 70\% of validation time in RPKI relying parties is spent on certificate parsing and signature verification, much of it unnecessary. Building on this insight, we introduce the improved RPKI (iRPKI), a backwards-compatible redesign that preserves all security guarantees while substantially reducing protocol overhead. iRPKI eliminates EE-certificates and ROA signatures, merges revocation and integrity objects, replaces verbose encodings with Protobuf, and restructures repository metadata for more efficient access.
We experimentally demonstrate that our implementation of iRPKI in the Routinator validator achieves a 20x speed-up of processing time, 18x improvement of bandwidth requirements and 8x reduction in cache memory footprint, 
while also eliminating classes of vulnerabilities that have led to at least 10 vulnerabilities in RPKI software. iRPKI significantly increases the feasibility of deploying RPKI at scale in the Internet, and especially in constrained environments. Our design may be deployed incrementally without impacting existing operations.

We make our design, object templates, publication point software and RP implementation open-source to facilitate integration of iRPKI into current RPKI deployments, and to enable reproduction of our study. We further provide recommendations how to derive new RPKI specification from our proposed improvements to facilitate standardization. 

\end{abstract}
\section{Introduction}
RPKI was designed to protect Border Gateway Protocol (BGP) from prefix hijacks~\cite{DBLP:journals/cacm/SunABVRCM21}.
Such hijacks can expose to espionage, theft of crypto-currency, outages, distribution of malware, and other devastating attacks~\cite{bgp:attacks,DBLP:journals/cacm/SunABVRCM21,klayswap, cloudfarehijack}.
To protect BGP, RPKI certifies the assigned network resources, and the routers validate the BGP announcements against them.

According to NIST monitor \cite{nistmonitor} as of April 2025, more than 55\% of the Internet prefixes are certified with RPKI and according to Rovista \cite{li2023rovista} almost 30\% of networks validate BGP announcements.
With the growing adoption of RPKI, it also plays an increasingly central role to the stability and resilience of the Internet.
Recently, RPKI also gained political importance: according to the routing security roadmap published by the US White House in 09/2024, RPKI is critical for national security~\cite{whitehouse2024routing}.
The roadmap identifies RPKI as a mature, ready-to-implement technology to mitigate vulnerabilities in BGP, and recommends deployment on all networks. It is thus expected RPKI deployment will continue to grow. 

RPKI was standardized relatively late compared to other protocols, like HTTPS \cite{rfc8446} or DNSSEC \cite{rfc9364}, allowing authors of the RPKI specification to learn from issues in other technologies and incorporate existing solutions into RPKI design. Observing that building a new technology from scratch is labor-intensive and error-prone,  RPKI utilizes many existing concepts from other protocols \cite{rfc6480}, like a Public Key Infrastructure \cite{rfc5280}, X.509 certificate templates and X.680 DER-encoding \cite{rfc8446}, and a distributed delegated server architecture similar to DNS \cite{rfc1035}. Designing RPKI on-top established technologies allows implementations to use existing tooling, like OpenSSL for parsing and validation, facilitating faster development and increasing acceptance by the community. 

However, RPKI is not merely a replica of other technologies; it aims to address the shortcomings of the fundaments it builds upon. For example, certificate revocation in TLS has historically proved ineffective with many TLS implementations not checking for revoked certificates at all due to the large client-side overhead \cite{liu2015end}. RPKI overcomes this issue by tightly integrating the Certificate Revocation List (CRL) in its architecture and mandating all clients to check it before validation \cite{rfc6487}. RPKI also acknowledges the issue of diverging certificate structure over time, resulting from the high flexibility for certificate issuers in TLS, and instead introduces multiple tight requirements to certificate structure \cite{rfc6487}. While these design decisions improve security, they make RPKI substantially complex, and make validation of certificates in RPKI more computationally intensive.

In this paper we investigate the architectural complexity of RPKI and analyze how its foundational design choices, such as reuse of legacy PKI structures, redundant cryptographic layers, and verbose data encodings, create systemic hurdles and challenges for  RPKI performance. We identify these issues across RPKI components, like validation complexity, correctness of implementations, resilience of RPKI deployments and future scalability. In addition, the complexity of the specification makes RPKI challenging to implement, leading to vulnerabilities \cite{van2022rpkiller, mirdita2023cure} and reducing the diversity of available RPKI implementations (only 1 open-source repository software and 3 RP implementations are widely deployed). Contrary to previous work, rather than attributing issues in RPKI to implementation flaws or insufficient resources \cite{van2022rpkiller,mirdita2024sok}, we show that many of the issues in RPKI stem from inherited complexity embedded in its object formats, certificate handling, and repository protocols. We perform a comprehensive analysis of RPKI Relying Party (RP) validators, correlating processing overhead with specific features mandated by the RFCs, and explore how the respective design choices result in intensive processing requirements for the software. We reveal that over half of validation time is consumed by non-essential certificate parsing and signature verification. Our evaluations on empirically derived datasets show that the complexity and inefficiency will further exacerbate as more RPKI objects are issued, more RPKI RPs are adopted and more use-cases, like ASPA, are incorporated, eventually leading to extremely resource-intensive and instable deployments and operation.

Based on these findings, we derive major improvements to the RPKI specification that reduce complexity, bandwidth requirements, object sizes and computational effort for implementations: We propose iRPKI, a redesign of RPKI that preserves the security and trust guarantees of the existing system while substantially simplifying its architecture. iRPKI introduces compact object formats, removes the need for per-object EE-certificates and redundant signatures, merges revocation and integrity mechanisms, and replaces inefficient encodings like XML and ASN.1 with lightweight alternatives such as Protobuf. We implement our improvements to RPKI design into the Routinator software package and experimentally show that our suggested changes not only reduce processing time by up to 93\%, and similarly bandwidth overhead by 95\%, cut the RRDP file sizes by 95\%, shrink cache memory usage by 88\%,  
but also eliminate entire classes of vulnerabilities rooted in parsing and validation logic, e.g., those related to EE-certificate parsing and redundant signature checks, that account for the majority of known RPKI implementation flaws \cite{mirdita2023cure}. 
We open-source iRPKI objects, the code of iRPKI based on the adapted Routinator code, and the repository tool\footnote{\href{https://www.dropbox.com/scl/fi/pvs6odh2ml25azifd6vg9/submission_ndss.zip?rlkey=eadvjuoxo7tar1q82h08cb4fu&st=elqnsfix&dl=0}{www.dropbox.com/scl/fi/pvs6odh2ml25azifd6vg9/submission\_ndss.zip}}.

{\bf Paper structure.} We review related work in Section \ref{sec:works}. We discuss RPKI in Section \ref{sec:background} and analyze its architecture, mapping the complexities in the current RPKI design and implementations, mapping them to RFCs, in Section \ref{sec:problems}. We analyze the effects of the complexity factors in the real world in Section \ref{sc:complexity} and illustrate how iRPKI overcomes them in Section \ref{sec:improvements}. We evaluate iRPKI in \ref{sec:evaluation} and discuss implications of our work in \ref{sec:discussion}. We conclude in Section \ref{sec:conclusion}.

\section{Related Work}\label{sec:works}
Previous research on RPKI explored vulnerabilities in RPKI implementations and errors in operation and proposed countermeasures \cite{morillo2021rov, su2024drr, hlavacek2023beyond, mirdita2024sok}. 
We next compare our work to previous research that developed mitigations and to papers that explored problems in RPKI, that we in this work trace to complexity in RPKI design.
\cite{hlavacek2023beyond} conducted an analysis of thresholds in RPKI software. 
Their work showed that the growing size of RPKI repositories leads to operational problems for RPs, but did not identify that the underlying cause for the large RPKI repository sizes were inefficiencies in the RPKI design. 
Our work proposes to fix the underlying factors that contribute to the growth of repositories beyond manageable limits, by improving the RPKI design and reducing the size of repositories by a factor of 18x, which enables smaller thresholds, as fetch times are decreased.

 \cite{mirdita2023cure} developed a fuzzing tool for RPKI software, and detected 18 vulnerabilities in RPs. The authors attributed the vulnerabilities to a lack of test tools for RPKI software developers, but did not identify the issues in the fundamental complexity of RPKI architecture. Our proposed design with iRPKI reduces the complexity of objects and RRDP, improving efficiency and reducing the susceptibility to bugs in parsing the complex RPKI objects. As we discuss in Section~\ref{sec:problems}, 10 of the vulnerabilities found by \cite{mirdita2023cure} would have been eliminated if iRPKI were used, since the improved object structures remove the fields and structure that triggered the vulnerability.

\cite{su2024drr} proposed a new hosting model for RPKI to overcome attacks and single points of failure of RPKI repositories. Their design additionally adds a continuous monitoring capability to RPKI, similar to Certificate Transparency (CT) in TLS. While their design enables detection of attacks, it does not improve the scalability of the RPKI architecture, as it increases the overall data amount in RPKI through federated storage. In contrast, our approach reduces the overall data in the system by improving objects and data exchange protocols directly. 
\section{Overview of RPKI}\label{sec:background}
The initial development of RPKI faced several design goals: How to attest which entity (Autonomous System) owns which network resources, how to authorize network resources in RPKI, how to distribute such ownership/authorization information securely, and how to validate the distributed information to allow routers informed decisions on BGP data. We explain how RPKI solves these design goals next.

\textbf{Ownership.} The design of RPKI follows the design of existing PKIs: Binding ownership of resources to a cryptographic key. In the context of RPKI, the ownership of IP prefixes and Autonomous System Numbers (ASNs) is bound to the ownership of a key through a signed certificate - similar to how HTTPS binds ownership of domains to a key through a signed TLS certificate. 
To validate ownership of resources, when certificates are issued and signed by a Certificate Authority (CA), RPKI follows the Internet number resource allocation hierarchy defined by IANA (top level), then Regional Internet Registries (RIRs) who allocate resources to entities like Internet Service Providers (ISPs), and so on - this hierarchy matches the way how resources are actually assigned in the Internet.  
The design of RPKI follows this delegation tree to establish trust: IANA anchors the global legitimacy of network resources delegation in RPKI. The five RIRs act as trust-roots and sign certificates for allocated resources. Owners of resources can further sub-allocate their resources, and use their key to sign the child certificate to attest this allocation. In RPKI, each certificate owner is a CA and can sign child certificates. Other entities can validate the ownership by following a trust chain, starting at the CA and validating each parent signature until validation reaches a trust root. 

\textbf{Authorizations.}
CAs can use their key to issue RPKI objects containing information for BGP routing. Prominently, RPKI standardizes Route Origin Authorizations (ROAs) that authorize a specific ASN to announce prefixes in BGP. CAs can only sign ROAs for prefixes they own. Similar to ROAs, RPKI also supports other authorization objects, like AS Parent Authorizations (ASPAs) attesting BGP providers, or BGPsec certificates holding a designated key for BGP path validation.

\textbf{Distribution.} 
The distribution of CA certificates is implemented through a hierarchical structure with RPKI repositories. CAs upload their certificate into the repository of their issuer, in most cases the repository of one of the five RIRs. RPKI also allows CAs to host their own repositories (distributed RPKI). While the CA certificate remains in the parent repository, the CA uploads all objects signed with its key to its own repository server. A link to this distributed repository is contained in the CA certificate, allowing discovery of the child server. A mixed model is also possible, where CAs keep their own private key but publish their objects through a hosted publication server.
In its current design, RPKI uses the repository infrastructure for CA certificates to also distribute all other objects. Thereby, all information is kept in a central and accessible location, allowing for easy and efficient batch downloads of all global RPKI data. Validation software starts at the root repositories and iterates through all CAs, following links to distributed repositories until all information from the global RPKI was downloaded.

{\bf Validation.} RPKI utilizes a middleware called RPs to fetch RPKI data from repositories. RPs access the global RPKI data, validate all objects, and compile a list of the validated data, called Validated ROA Payloads (VRPs). Downloading data from repositories is implemented through two access protocols. Initially, rsync was used to fetch data from remote repositories. Eventually, a second transport mechanism, RPKI Repository Delta Protocol (RRDP), was standardized \cite{rfc8182}. RRDP allows incremental fetches of repository data by including incremental changes in Delta.xml files. The RP software is informed of the current repository state through a serial number and session id inside a Notification.xml file that is linked in the certificate of all CAs publishing at a given repository. If an update over deltas is possible, the deltas are applied to the local state incrementally until the current serial number is reached. If the delta update fails or no local state is available, e.g., in the first start-up of an RP, the entire repository state is downloaded through a Snapshot.xml file. The snapshot contains the names and base64-encoded binary content of all objects in the repository. 
Routers download the VRPs from RPs through the RPKI to Router protocol (RTR) and use it to validate incoming BGP messages.

\textbf{Integrity and revocation.}
Signatures on ROAs ensure authenticity, but not repository integrity, e.g., a deleted ROA would not be detected. 
RPKI includes an additional object for each CA, called the manifest, to ensure integrity of the content in the repositories. Manifests are signed by the CA and contain a list of names and hashes of all objects issued by the CA. Through the manifest, RPs validate that the repository content is complete and unaltered. A successful validation of a repository requires a valid manifest.  
Removing an object from the manifest does not equal revocation. Instead, RPKI repositories additionally include a Certificate Revocation List (CRL) to remove unexpired object that should be invalidated. The CRL lists the serial numbers of revoked objects.

\section{Complexity of RPKI}\label{sc:complexity}
The design of RPKI reflects a deliberate choice to build on existing PKI technologies. While this approach has the benefit of leveraging mature standards, it also introduces significant complexity to RPKI. In this section we explore the impact of adaptation of generic cryptographic and transport technologies to RPKI. We show how they undermine the goal of resilience, effectively resulting in a complex, fragile system with unexpected operational challenges. By analyzing RPKI design, we explore the factors contributing to its complexity. 

Our analysis additionally uses current RPKI software, identifying which components lead to the largest processing overhead. We select the most popular RP implementation Routinator for our analysis, which also proved the fastest in our test. 
The motivation behind using Routinator is clarity of our results: All RPs generally follow the RFC requirements for their processing, and thus implement the same core logic. Routinator has the fastest processing times, illustrating it has lowest non-RFC overhead and thus the lowest risk of skewed results from non-RFC related functionality, like internal state construction, key management, or cache modifications. 

\subsection{Analysis Methodology} 
For our RP analysis, we use the Linux perf flamegraph tool for Rust. 
It allows us to quantify the overall fraction of processing time Routinator spends in different parts of the software. We compare the flamegraph with those for other RP implementations to ensure our findings are comparable and find similar results on the distribution of processing time spent in the different core modules, like parsing and validation. 

To ensure reproducibility of our results and reduce the outside noise of our evaluations, we create a local, isolated setup, and run our analysis in it. This is important, as real-world processing times are heavily influenced by external factors, like configuration errors or networking issues in remote repositories \cite{mirdita2024sok}.
For our setup, we create a local RPKI repository with 100 test CAs. We choose 100, since we experimentally find that the flamegraph results do not change when including more CAs. Each CA includes one manifest and one CRL. For each CA, we additionally include 6 ROAs, the average amount of ROAs per CA in real-world RPKI. To serve our local repository, we use an nginx webserver without bandwidth limitations. Since download times relate to object size, looking at the size of objects allows us to quantify impact on download times, omitting the need to limit download speed for analysis. 
We explain how our results generalize to real-world RPKI deployments, including limited download speed and large amount of CAs in Section \ref{sec:evaluation}.

\begin{figure}[t!]
    \centering
    \includegraphics[width=0.95\columnwidth]{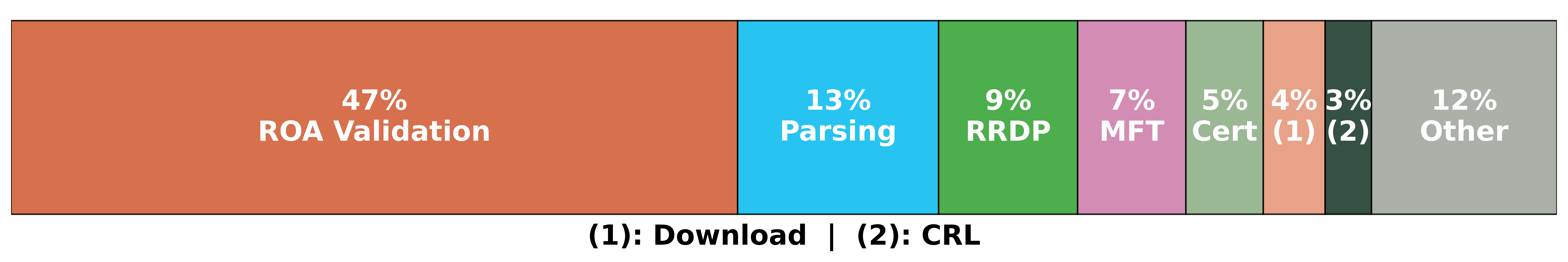}
    \vspace{-5pt}
    \caption{Flamegraph Routinator 600 ROAs.}
    \vspace{-15pt}
    \label{fig:bar_ex_plane}
\end{figure}

\textbf{Flamegraph results.}
The results of our evaluation with Routinator RP are shown in Figure \ref{fig:bar_ex_plane}. 
{\updated The majority of processing time is spent to validate the signed objects, totaling to 62\% of RP processing.} This processing time is spent to validate 1200 signature on ROAs (2 per object), 200 signatures on manifests (2 per objects), and 100 signatures on CRLs and CA certificates (1 per object). Both ROAs and manifests contain 2 signatures per object since they contain a certificate and a signed object signature. For example, looking at ROAs (47\% of total processing time) Routinator spends 24\% of total processing time on validating the certificate signature and 23\% to validate the ROA content signature. However, not all processing time is spend on public key validation. 13\% of processing time is spent on parsing and validating non-signature fields of signed objects (ROAs and MFTs). Further, parsing and validating the RRDP XML files takes 9\%, with the large majority of the time spent on parsing the Snapshot. 
HTTPS request handling, like socket creation, takes about 4\% of total processing time. Finally, about 12\% of processing time is spent in other functions, like internal state management, thread management, and disc writes. In conclusion, the RP spends the large majority of time on cryptographic validation, followed by parsing of objects and snapshot files. 

Next, based on the results of our evaluation, we analyze the components in the design of RPKI that consume the resources and derive the factors contributing to the complexity of RPKI.

\subsection{Complexity of EE-Certificates}
The structure of RPKI objects makes heavy use of existing templates. CA certificates and CRLs use X.509 certificate templates, defined with ASN.1 syntax and encoded with Distinguished Encoding Rules (DER), corresponding to how HTTPS handles certificates and CRLs. All other objects in RPKI, like ROAs and manifests, use the signed object template from Cryptographic Message Syntax (CMS). Specifically, they all contain signed object content, like the ASN and IP prefixes for ROAs, a signed certificate, and a signer information field containing a signature over the signed content. The certificate in these objects is not the CA certificate, but an individual one-off End-Entity (EE) X.509 certificate issued for each object. The EE-certificate contains a one-off key that is only used to sign and verify the content of the object where the certificate is included. Since all objects are signed twice, validating RPKI signed objects also requires two signature validations; first, the RP validates that the EE-certificate was validly signed by the respective CA. Then, the RP uses the public key in the EE-certificate to validate to signature over the object content.

This design is supported by the ASN.1 type \textit{signedObject} \cite{rfc5652}, allowing to add a signature on encapsulated content, and optionally include signer certificates. Adding the signing certificate directly to objects is mainly motivated by self-sufficiency: According to \cite{rfc5652} Section~5.1, this design allows the object to contain \emph{all} certificates necessary to validate it up to a trust-root, making it fully self-sufficient in validation. The RFC acknowledges that adding the full chain is not required, and objects may contain only a subset of certificates since other necessary certificates for path validation can also be attained through other means. The latter option is used in RPKI: RPKI objects include a single certificate that is only used to validate this specific object content. All other certificates for path validation up to the trust root are obtained through the RPKI architecture and are not contained within the signed object. 

This hybrid design (only containing a subset of required certificates) raises the question why RPKI includes any certificates in objects instead of using the other validation option given by \cite{rfc5652}, which is omitting any contained certificates and having the content directly signed by the CA.

While not explicitly stated as the reason in the RFC, this design is likely based on supporting potential future use-cases. Including EE-certificates in each object allows more independent distribution of signed objects, as it makes objects partially self-sufficient: The RP can use certificate fields for content validation, and to find issuer certificates for path validation. This potential use-case is specifically mentioned in \cite{rfc6480} §3.1, describing ROAs might be distributed, e.g., within BGP messages. However, to-date, out-of-band distribution of ROAs is neither supported by the specification, nor is the creation or processing of such objects supported by any current repository or RP implementation. To the best of our knowledge, no current efforts within the IETF or software development community are working on such a use-case.

Including the EE-certificate also provides benefits in the current architecture. \cite{rfc6487} states that EE-certificates allow targeted revocation of a signed object through its certificate.
By using EE-certificates, revocation of objects can use the serial number of the certificate within the signed object instead of having to revoke the CA certificate, re-issuing it and re-signing all child objects. It improves revocation efficiency.
Using EE-certificates has the additional advantage of re-usability of existing technologies and open-source solutions \cite{rfc6488}. For example, attesting additional aspects of an object (like issuer name) is implemented through existing certificate extensions instead of developing new concepts for storage within object content. By using EE-certificates, RPKI thus aims to lower the complexity of implementing new object templates by using existing (complex) template definitions where possible.

\subsection{MFT and CRL}
Every CA in RPKI requires a signed manifest and a signed CRL to ensure integrity and track revocation of repository content. This sets RPKI apart from other technologies like HTTPS, which only includes CRLs and therefore has smaller per-CA overhead (1 object instead of 2), raising the question why RPKI increases complexity by using two such objects. 
\cite{rfc6486} states that manifests are modeled after CRLs, sharing many characteristics, like their fundamental structure and their use of serial numbers to assure a consistent sequence between new issuance of CRL / MFT respectively. 
Further, both serve similar roles in the architecture: Ensuring validity of objects from the perspective of the CA. Their combined role is acknowledged in the RFCs, stating that manifests and CRLs should always be re-issued together.
This advice is generally followed by operators, and we find all production CAs re-issue their CRL if they re-issue the manifest, and vice versa.
Any change in the repository, like a new or deleted object or a refresh of validity will thus always lead to re-issuance of manifest and CRL. 
Despite their similar function and even though they always must be changed together, manifest and CRLs are separate, individually signed objects. Keeping the objects separate creates overhead for issuance, downloading, cache size, and validation effort. 

While the RFC does not provide insights into the motivation of having two individual objects, separating CRL and manifest is likely rooted in their slightly different roles, the manifest handles integrity and the CRL handles revocation. 
However, the benefit of this logical separation is limited. It does not ease issuance of objects, as a change to revocation or repository content always requires both objects to be re-issued. Checking CA-side validity of an object, i.e., if a validly signed object might be excluded from the manifest or is revoked in the CRL, is also not made easier, as RPs must always consult both objects before deciding any object is valid, making validation logic and debugging more complex.

\subsection{RRDP File Structure}
RPKI uses RRDP as its primary data exchange protocol, which utilizes XML-encoding for all files. The RFC does not give reasoning \textit{why} XML was chosen, however, using XML likely follows the design philosophy of using existing and accepted technologies in RPKI \cite{rfc6480}.

XML comes with significant overhead. It is designed for human-readability, not maximum efficiency and thus has larger file-sizes and processing times than other (binary-) protocols \cite{gligoric2011performance}. Further, XML offers significant flexibility in object structure and many features for processing logic, at the cost of parsing effort and validation requirements to protect against XML-attacks (like million laughs attack or xml-bomb) \cite{singh2019taxonomy}.

Since RRDP nests the entire repository content inside XML files, XML files have substantial impact on download, parsing, and processing times of RPs.

\subsection{ASN.1 Encodings}
All RPKI objects use ASN.1 object templates, and are encoded / decoded with DER binary encoding. This design is consistent with the design of other internet protocols like TLS, allowing RPKI implementations to use existing open-source tooling for parsing.

While this design is sensible for CA certificates, which use X.509 formatting like TLS and may heavily re-use existing implementations for parsing / validation, the benefit from ASN.1 is much lower for other RPKI object types like ROAs that require new validation implementations anyway. DER encoding/decoding is complex and difficult to implement correctly \cite{mirdita2023cure}, leading to vulnerabilities. Further, our experiments show that encoding objects with protobuf improves processing speed in RPs, as we illustrate in Section~\ref{sec:evaluation}.

\section{Complexity Creates Hurdles for RPKI}\label{sec:problems}
The complexity of RPKI design and implementations come with a price and already cause issues in practical operation. We show that these issues will further exacerbate with the growth of RPKI deployment.

\subsection{Bandwidth}
Currently, 91 repositories and 4211 RPs are deployed in production RPKI systems, measured on April 17 2025. 
Each of these repository servers must offer sufficient bandwidth to accommodate requests by all RPs, which regularly poll all servers. The exact required bandwidth is a function of the number of RPs that download the data, the size of the files within the repository, and whether RPs download deltas or snapshots. 
In normal operation, most RPs download deltas, which are generally smaller than snapshots \cite{rfc8182}. But cold starts, session refreshes, or faulty deltas lead to downloading complete snapshots, and the repository must ensure bandwidth is sufficient for snapshot downloads by all RPs \cite{bcpserver}. 

Mapping the current global RPKI state, the average RRDP snapshot size is around 15 MB. 
Combining this size with data on RP distribution \cite{rovnlnetlabs}, we can approximate the required minimal bandwidth of repositories to around 670 Mbit/s. An interested reader is referred to Appendix~\ref{app:poll_times} for detailed computations. This bandwidth estimate provides a lower bound, not accounting for HTTPS overhead, retransmissions, and RP fallbacks. 
In practice, repositories need to provide more than 670 Mbit/s bandwidth to ensure sufficient resilience against network errors and attacks, requiring repositories to host their servers on high-availability, high bandwidth infrastructure.

Since RPKI is continuously developing and new use-cases are still being standardized, the amount of RPs and the amount of data in repositories is growing \cite{mailingListReview}. {\updated Currently, ca. 55\% of globally announced address space is covered by RPKI ROAs
and measurements using representative samples show around 28\% of ASes and 20\% of users are protected by ROV enforcement using an RP \cite{li2023rovista, hlavacek2023keep, apnicrov}. 
Extrapolating RPKI deployment to 100\% ROA and 100\% ROV coverage - assuming linear ROA growth and a stable ratio of RPs per enforcing AS - yields a doubling of average snapshot size to 30MB, and a 5x increase in the amount of RPs to 21,055.} Thus, in a conservative approximation on full RPKI deployment 
average repositories will require minimal bandwidth of 6.7 GBit/s. For larger repositories, the required data-rate is even worse. Applying the same approximation to the RIPE NCC snapshot, which is currently 230MB, RIPE will have to provide at least 102 GBit/s bandwidth. The bandwidth requirements of current RPKI thus do not scale well for growing adoption of RPKI, even under best-case assumptions on use-cases, congestion, and other networking errors. 
More objects and snapshots additionally increase the fetch durations of RPs, exasperating issues with low flexibility and resilience. {\updated While longer default RP fetch intervals could lighten load on repositories, long intervals are disadvantages since they extend propagation delay between the upload of new objects and impact in routing. For example, if a ROA is uploaded but fetch intervals are long, new BGP announcements can conflict the old ROAs in all RPs that have not yet updated their cache, leading to invalidation of BGP announcements and thereby loss of traffic. The longest current default interval of production RPs is 1h. }

Standardization of new use-cases will also increase the amount of RPKI objects and introduce complexity to the RPKI architecture, as each new object type requires new parsing and validation logic in implementations. For example, if ASPAs reach the same deployment level as ROAs, it will increase overall RPKI size by roughly 41\%, and will require all implementations to support ASPA validation.

\subsection{High Latency Fetch Intervals}
Setting up current RPs and running a full fetch in a default configuration takes 7min on Routinator, 7.1min on rpki-client, and 10min in Fort, with between 71\% (Routinator) and 58\% (Fort) of time spent on downloading RPKI data, and 29\% (42\%) of time spent on processing / validation of data\footnote{We run all RPs sequentially on a machine with 100 Mbit/s available bandwidth and 8-core i7 processor.}. 
{\updated Investigating the RPs, we find that if fetch times increase beyond download thresholds, it will lead to data loss;
we provide more detailed considerations on threshold limits and object expiration in Appendix~\ref{app:fetch_issues}. 
Full RPKI adoption will increase RP fetch times, increasing risks of exceeding the thresholds.  Contributing to this issue, the number of repositories is continuously growing \cite{hlavacek2023beyond}, which is extending RP fetch times further. Multiple previous research works raised issues with scalability of the architecture with growing numbers of repositories \cite{su2024drr, hlavacek2023beyond}, as RPs have to contact each repository individually, and configuration errors in the growing number of repositories further extend fetches.}

Furthermore, fetches include downloading of around 1.2 GB of data, and requires validation of around 764000 RSA signatures, also contributing to high latency fetch intervals if resources are insufficient \cite{mailingListLatency}. To illustrate how constrained environments can impact RPs, we repeat the RP measurement with bandwidth constrained to 10 Mbit/s. This increases total validation time 3x, with Routinator and rpki-client taking around 21min to finish and Fort taking 24min.
We provide more detailed considerations on the implications of long fetches in Appendix~\ref{app:long_fetches}.

\subsection{Insecurity Through Complexity}
Complex and intricate protocol requirements make secure and correct implementation of the specification challenging. In RPKI, a substantial amount of vulnerabilities was discovered over recent years that can be mapped to complexity. 

Looking at all CVEs for the actively maintained RPs and all RPKI papers published in the last five years, we find a total of 47 vulnerabilities. The majority (29/47) of these vulnerabilities stem from errors in the decoding and processing of RPKI objects. This is not surprising, as the decoding and processing of RPKI objects incurs substantial complexity, including DER decoding of at least six different currently standardized object types, and each RPKI object including interconnected validation steps. 7 vulnerabilities in RPKI were discovered in the RPKI-specific data exchange protocol RRDP, including issues with compression, XML-specific attacks, content decoding, and path-traversal \cite{van2022rpkiller, mirdita2022poster, mirdita2023cure}.

Complex protocol requirements also lead to inconsistent implementations. Previous work \cite{mirdita2023cure} illustrated seven cases where RP implementations reached differing results on the validity of RPKI objects. This is problematic in the context of RPKI, as supposedly valid objects that fail validation in a subset of RPs will lead to these objects being excluded from the VRPs output of the RP, and thus unprotected from BGP hijacks. Since the RPs react inconsistently, the issuer might not even notice the error, if the object is not tested with all available RP implementations before publication.

\subsection{Algorithm Agility in RPKI}
Currently, RPKI solely supports RSA-2048 for its public key signatures \cite{rfc6485}, but the need to transition to new algorithms is acknowledged in RFC6916. This transition will be required for supporting quantum-secure algorithms, a shift the security community aims to complete roughly until 2035 \footnote{\href{https://www.europol.europa.eu/media-press/newsroom/news/call-for-action-urgent-plan-needed-to-transition-to-post-quantum-cryptography-together}{www.europol.europa.eu/}, \href{https://www.ncsc.gov.uk/guidance/pqc-migration-timelines}{www.ncsc.gov.uk/}, \href{https://nvlpubs.nist.gov/nistpubs/ir/2024/NIST.IR.8547.ipd.pdf}{nvlpubs.nist.gov/}}. Depending on which algorithm will be implemented into RPKI, the overall size of the downloaded RPKI data will increase from today 1.2 GB to between 2.9 GB and 39.1 GB. For more details on size computations, see Appendix~\ref{app:postquant}. As all proposed post-quantum algorithms increase overall object size, transitioning to a post-quantum secure algorithm will thus put additional load on the RPKI architecture.

\section{Improved RPKI}\label{sec:improvements}
The deployment and scalability issues in RPKI stem from complexity and inefficiency of the architecture. This is also evident in the RFCs; while security considerations are included in the majority of RPKI RFCs, the specification lacks discussions on the efficiency of design choices. Doing such efficiency considerations for different aspects of RPKI is not trivial:
The substantial complexity and inter-dependency of RPKI specification makes quantifying the real-world impact of different design choices challenging. For example, changing one component to increase efficiency can negatively impact other inter-dependent components, negating the postulated benefit. This can even happen with improvements proposed through RFCs, like RFC8360 proposing improvements to the path validation algorithm which was never deployed, likely because the proposed algorithm only provides minor improvements to path construction while breaking object compatibility with legacy RPs. Evaluating the implications of design choices is made more complicated by the intricate security architecture of RPKI, as changing any mechanism may break RPKI security. 

For example, deducing the overhead that validating EE-certificates introduces is hard from RFC text alone, as EE-certificate validation includes 15 different fields, two signatures, and certificate path validation with other certificates. Changes to the EE-certificate, like omitting a field, can break real-world use-cases or, worse, enable attacks. Evaluating the impact of design choices thus requires a practical comparison between the current design and alternative approaches that provide identical use-cases and security. For the EE-certificate, this can, e.g., include evaluating the real-world processing time and security benefits of EE-certificate validation compared to other mechanisms to establish authenticity of object content. 

Our methodology uses practical insights and direct comparisons to alternative approaches to evaluate real-world impact of design choices in RPKI, and derive improvements to these designs. 
For this, we first define necessary design goals that all changes to RPKI need to fulfill to ensure changes are meaningful and as secure as current RPKI design. 
We then use the practical RP study from Section~\ref{sec:background} to investigate the components of RPKI that result in the largest overhead regarding object size, processing time, and cache footprint. We analyze the RFC design choices underlying the components, identify overheads, and derive changes to the design of RPKI that overcome the inefficiencies to significantly improve performance. 
We contextualize all changes within the RFCs to ensure new designs do not break current use-cases, lower security, or introduce other undesired side-effects in the interaction with other components of RPKI. Finally, we implement all changes into RPKI software and evaluate them against the current design to show the improved design outperforms current RPKI in practice.

\subsection{Design Goals}
Our {\updated conceptual} design goals ensure that the improvements to RPKI are secure, deployable, and maintainable. 

\textbf{\#1 Security.} 
RPKI is a security critical protocol used to defend against practical attacks, like prefix hijacks. Any changes made to the protocol or implementations must thus not lower the security guarantees of RPKI. 
Specifically, changes must not impact the security of path validation (i.e., validating the authorization of a given entity to issue specific objects), and may not lower the security of objects (i.e., validating if a given objects was issues by a specific entity). 
Importantly, the design goal does not mitigate changes to \textit{how} these security requirements are achieved, as long as changes provide an equal level of protection and do not enable attacks.

\textbf{\#2 Backwards compatibility.} 
Disruptive breaking changes to RPKI are challenging to implement, as RPKI is widely deployed and runs in production systems. 
Any changes to the RPKI must therefore provide backwards-compatibility: Deploying the new changes must seamlessly integrate into current RPKI operation. Specifically, deployments of current (legacy) and improved RPKI must be able to operate side-by-side without impacting each other, and an incremental deployment of the changes must be supported. New software versions should not incur any manual effort by operators; all changes should be transparent to users of the software. Changes should further not break any existing use-cases.
This makes deployment significantly easier, as no re-trainings or adaptions to setups besides software updates are  necessary. 

\textbf{\#3 Re-use existing solutions.} 
RPKI heavily utilizes existing concepts and solutions, like ASN.1 for encoding and X.509 for certificates, a design goal specifically mentioned in \cite{rfc6480}. This is sensible to ensure minimal development overhead for implementations and improve acceptance by the community. {\updated However, re-using existing solutions incurs a trade-of between resulting overhead and gained benefits. Any changes to the design should thus weigh benefits against downsides. For this, we evaluate each proposed change for its complexity and performance impact. Where changes do not incur substantial benefit, the improved design re-uses the current RPKI structure, and generally keeps changes as minimal as possible.}

\textbf{\#4 Major Improvements.} 
Since RPKI is already deployed, there is a natural aversion against changes to the protocol that may cause disruptions or other issues in production systems. 
Thus, all proposed adaptions must provide major and tangible improvements to RPKI operators 
to warrant the inherent risk of changes to production systems. 
Specifically, improvements should significantly lower processing times, download times, or reduce memory footprint. 
To quantify what constitutes a major change, we run our own RPKI RP software and observe natural fluctuations of the above described metrics. 
We find natural fluctuations of up to 5\% between consecutive runs of the RPs\footnote{Download times fluctuate up to 50\% due to changes in repository reachability. On consecutive runs without changes in reachability of remote repositories, the 5\% observation holds.}. 
We define an improvement as significant if it improves one of the above metric by more than double its natural fluctuations, i.e., improves by at least 10\%. All changes we make in iRPKI fulfill this requirement. 

Based on these design goals, we derive the following improvements to RPKI specification.

\subsection{Removing the EE-Certificate} 
\begin{figure}
    \centering
    \includegraphics[width=1.0\linewidth]{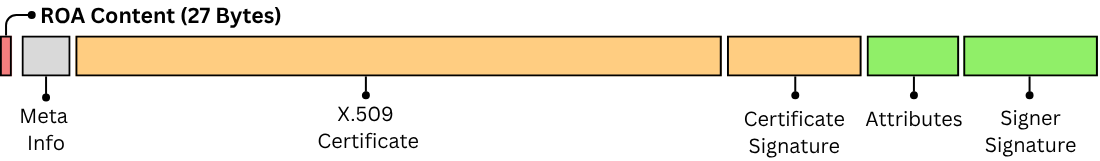}
    \caption{{\updated ROA object structure in current RPKI.}}
    \label{fig:roa_struc}
    \vspace{-10pt}
\end{figure}

The current design of RPKI uses EE-certificates to establish authenticity of signed objects, like ROAs. 
Using EE-certificates instead of having CAs directly sign object content has substantial impact on object size and validation effort. Figure \ref{fig:roa_struc} illustrates the real-world component sizes in a best-practice ROA for a /24 IPv4 prefix. Over 72\% of the object is taken up by the EE-certificate and its signature (1337 / 1846 Bytes), while the actual ROA content, i.e., ASN and authorized IP prefix, only make up 1.4\% (27 / 1846 Bytes). This is a substantial overhead, with almost 3/4 of the object taken by data required for validation, not by content. Further, looking at the flamegraph of Routinator from Section~\ref{sec:background}, roughly 52\% of total signed object validation time is used to parse and validate the certificate, its fields, and key. 

\textbf{Do EE-certificates improve security?} 
Using EE-certificates in RPKI does not provide stronger security than signing the object with the CA key, as both use the same signing procedure and cryptographic cipher. The support for built-in EE-certificates is motivated by (validation-) convenience and object self-sufficiency, not added security \cite{rfc5652}. Adapting the design to change or remove the EE-certificate is, however, not trivial. 
The current RPKI design relies on EE-certificates for validation, for example to contain the signing key and validity period of an object. Further, EE-certificates also contain multiple fields essential for RP processing, like the parent key fingerprint or owner name. Making changes to EE-certificates requires careful considerations on how each EE-certificate field is used in object validation. 

\textbf{EE-certificate fields.}
Within the EE-certificate, 10 fields are necessary for the validation of the certificate itself, not the content it authenticates. These fields contain the certificate version, map the certificate to its respective issuer, identify the owner of the certificate and their key, how the certificate may be used, list the signature algorithm, describe where the CRL of the certificate can be found, where the issuer certificate is located, where the object is located, and which policies should be applied when validating this certificate.

All of these fields are only used for certificate validation, and are not used after certificate validation finished.
Some fields of the certificate are, however, critical to the validation and security of the content signed with the certificate key.

\textbf{Validity.} The validity period of an RPKI object is attested through its EE-certificate: If the certificate expires, the object becomes invalid. Removing the certificate takes away the possibility to check expiration through its validity field. EE-certificates are, however, not the only means to check expiration in RPKI. The manifest additionally includes a validity period within its content. Trivially, such validity periods can be added to all other signed objects, omitting the need to verify expiration solely through the EE-certificate.

\textbf{Revocation.}
In RPKI, signed objects are revoked by adding their certificate serial number to the CRL. This design, which is aligned with revocation mechanisms in other technologies like TLS, requires that each object contains a unique serial number, a requirement that is currently implemented through certificate serial numbers. Serial numbers are, however, not limited to certificates. 
Both manifests and CRLs contain additional serial numbers in their content (MFT number / CRL number).
Following the manifest and CRL templates, a serial number can also be added to all other signed objects to allow targeted revocation, omitting the need for revocation solely through certificate serial numbers. 

\textbf{Signing key.}
The EE-certificate contains a one-off signing key for validation of the signature over the object content. 
In an alternative design, the CA directly signs the content of the object, instead of signing the certificate which is then used to sign the content. This is feasible, e.g., the CRL of a CA is directly signed by the CA. 
The downside of removing the one-off key is that the CA can no longer revoke individual keys to revoke objects. However, by including the serial number in the object directly, the CA can simply revoke the object instead of revoking the key on the object. 

\textbf{Delegations.}
EE-certificates contain the IpAddrDelegation and ASNumberDelegation extensions to attest resources allocated to this certificate. The validation of the IP and ASN resources is conducted through the CA certificate. 
The RFC does not make clear \emph{why} these delegations are included in EE-certificates and in manifests, both extensions must be set to null (inherit) \cite{rfc6486}. ROAs need to have explicit values in the extension according to the RFC and RPs must validate them. 
However, ROA validation does not make use of the values in the ASN delegation, and while it validates that the ROA resources are contained in the IP resources of the EE-certificate, these resources must be a subset of the CA resources anyway, so there is no clear benefit of additionally including IP address delegation validation in the EE-certificate, after it was already validated through the CA certificate. 
Our tests show that both fields could be omitted or set to inherit in all signed objects without reducing security / authenticity, as validation of all RPs will fail if these resources conflict with the resources stated in the CA certificate. 

\textbf{Removing the EE-certificate.}
Following our above analysis, none of the signed object validation steps can only be implemented through the costly EE-certificate. The 10 fields that only serve certificate validation are not required if no EE-certificate is used, and all other necessary fields implement functionality that can also be implemented through fields within object content. By adapting the design to omit EE-certificates, the efficiency and object size can be improved. This new design requires some adaptions to object templates:

Without the EE-certificate, no one-off key can be added to the object. Instead of signing the EE-certificate and using the one-off key to sign the object content, CAs sign the content of the object directly, like for CRLs. 

Parsing signed objects without a certificate is already supported by existing software, like OpenSSL, as included certificates are optional as per CMS \cite{rfc5652}. Investigating the implementations, we find that Routinator has a strict custom implementation of signed object parsing, and requires minor code adaptions ($\approx$ 20 Lines of Code) to remove the need for the existence of an EE-certificate in parsing.

As expiration is checked through certificate validity, a new validity field is required in all signed objects expect the manifest, which already includes a validity field. Following manifest structure, the validity can be included as two UTC-time ASN.1 fields. 

The revocation mechanism requires shifting from certificate serial numbers to content serial numbers. For manifest and CRL, the existing content serial numbers can be used for this, as they already follow the same requirements as certificate serial numbers. For all other signed object, a unique serial number needs to be added to the content, like a ROA number. 

Implementing these changes to the signed object template allows significantly smaller objects, and halves the amount of required signature validations. 
We illustrate the performance and size improvements of removing the EE-certificate in different deployment scenarios in Section~\ref{sec:evaluation}.

\subsection{Removing Signatures on ROAs.}
{\updated Signatures on RPKI objects ensure authenticity, as RPs can validate the signature of the CA to confirm that the signed object was validly issued. Signatures do not protect integrity:} A malicious repository may still delete some objects issued by the CA, compromising integrity and availability. To detect such attacks, RPKI uses the manifest file. 
RPs compare downloaded repository content against the manifest and fail validation of the entire CA if manifest validation fails, as the failed validation indicates an attack on repository integrity, like deletion of an object. 
The manifest, however, not only detects integrity compromise but additionally offers a signed protection of authenticity: If an attacker changes any object covered by the manifest, like a ROA, the hash of the tampered file will differ from the hash in the {\updated manifest. This leads manifest validation to fail, invalidating the entire CA and preventing acceptance of tampered objects.} Since the manifest is signed by the CA, the attacker can not manipulate the manifest hash to the new malicious object content to circumvent validation failure. 
While the initial design goal of manifests was integrity, they thus offer a redundant mechanism to verify object authenticity. In practice, if an attacker manipulates a ROA, breaking authenticity, the validation will not fail when the RP validates the ROA signature, but when the RP validates its manifest hash. ROA signature validation and manifest hashes both detect compromised object authenticity.

This authenticity protection of manifests is possible in a threat model of an external attacker, i.e., an attacker without ownership of the private key \cite{rfc6486}. If the attacker has ownership of the CA private key, they can adapt the ROA hash on the manifest content. However, in this case, the ROA signature would also not protect the CA, as the attacker can forge valid ROA signatures using the CA key.

In the current design of RPKI, signatures on ROAs merely add an \textit{additional} means to detect tampering of the object, but do not improve overall security. ROA authenticity is validated through the manifest, and signature validation of ROAs will only fail for mistakes by the CA, not in malicious attacks. 
At the same time, since ROA are the most abundant objects in the RPKI (76\% of all RPKI objects are ROAs), validating their signatures takes up a major part of total RP processing time. 
Removing signatures from ROAs does not reduce security while providing substantial benefits in lower validation load on RPs, and smaller size of ROA files.

\textbf{ROA without signature.} 
Removing ROA signatures does not require adaptions of manifest content, but modifies its stated role in \cite{rfc6486} from integrity protection to additional authenticity protection. 
This has implications on the cryptographic and availability requirements for the manifest.
Currently, manifests use sha256 for hash validation, which is considered secure for cryptographic applications \cite{gilbert2003security}. 
Removing ROA signatures requires strict availability of manifests, as ROAs can not be validated without it. This requirement is already part of current RPKI, and all RPs will fail validation of any repository if the manifest is unavailable. 
RPs will not accept any ROAs that are not listed on the manifest or have invalid manifest hashes, even if they are validly signed. 
The new role of the manifest thus does not lead to any changes in threat model or availability requirements, and RPs do not need to implement any new checks.
We discuss the implications of removing ROA signatures for potential future use-cases of RPKI in Section~\ref{sec:discussion}.

\subsection{Improving ROA Structure}
The current template for ROAs uses the CMS signedData type, supporting signatures and certificates within the file. After removing certificate and signature from the ROA, the type no longer fits the ROA use-case. 
Instead of nesting the ROA econtent inside an outer object structure like CMS, the ROA can now be reduced to its actual content, i.e. serial, validity, ASN and IP prefixes, reducing the final size of a ROA from 2130 bytes (with signature and EE-certificate) to 80 bytes. We provide the updated ASN.1 template in Appendix~\ref{app:template_asn1_roa}.

\subsection{Combining CRL and MFT}
Current design of RPKI has CRLs and manifests as two separate objects, despite their similar roles and structures, shown in Section~\ref{sec:background}. The similarity of CRL and manifests and benefits in combining them are evident in the implementations of Routinator and rpki-client, both integrate CRL and manifest into one unified structure for internal processing.

Changing the design of RPKI to combine CRLs and manifests into a single object provides benefits: 
A combined object not only makes processing easier and more efficient, but also improves clarity of repository structure. Instead of consulting the manifest and the CRL for CA-side validity of an object, the RP / user must only check a single file that contains all integrity and revocation information for all objects by a CA.

To implement this new combined object, we define a new ASN.1 structure for manifests that integrates the CRL, illustrated in Appendix~\ref{app:template_asn1_mft}. The template is based on the manifest and adds the CRL field \textit{revoked certificates} after \textit{manifest hashes}. All other necessary information contained in a CRL is already present in the manifest, the CRL parentFingerprint extension is identical to the manifest issuerID, and the manifest number serves the same purpose as the CRL number extension. 

RP validation of the manifest is adapted to additionally include checking if any object in the repository is revoked through the revoked serials field of the manifest. When this new manifest is used, a standalone CRL is not required and must be omitted by the issuer / ignored by the RP.

\subsection{Improved RRDP File Structure and Encoding}
We identify several points where the structure of RRDP files can be optimized to improve efficiency. 
First, each entry in the Snapshot XML contains the respective file name. For example, consider the first entry in the AFRINIC XML snapshot {\footnotesize rsync://rpki.afrinic.net/repository/04E8B0D80F4D11E0B657D8931367AE7D/\\62gPOPXWxxu0sQa4vQZYUBLaMbY.mft}. The file URI contains an individual file name 62g[...].mft, and the repository URI of the respective CA (rsync://[...]). Including the full repository URI, 69 bytes, in each entry leads to an average overhead of around 2\% per entry. The relative overhead, however, increases for smaller objects. For improved ROAs with 80 bytes of content, the full URI makes up 46\% of the XML entry. We provide more in-depth evaluations in Section~\ref{sec:evaluation}.

\begin{figure}
    \centering
    \includegraphics[width=1.0\linewidth]{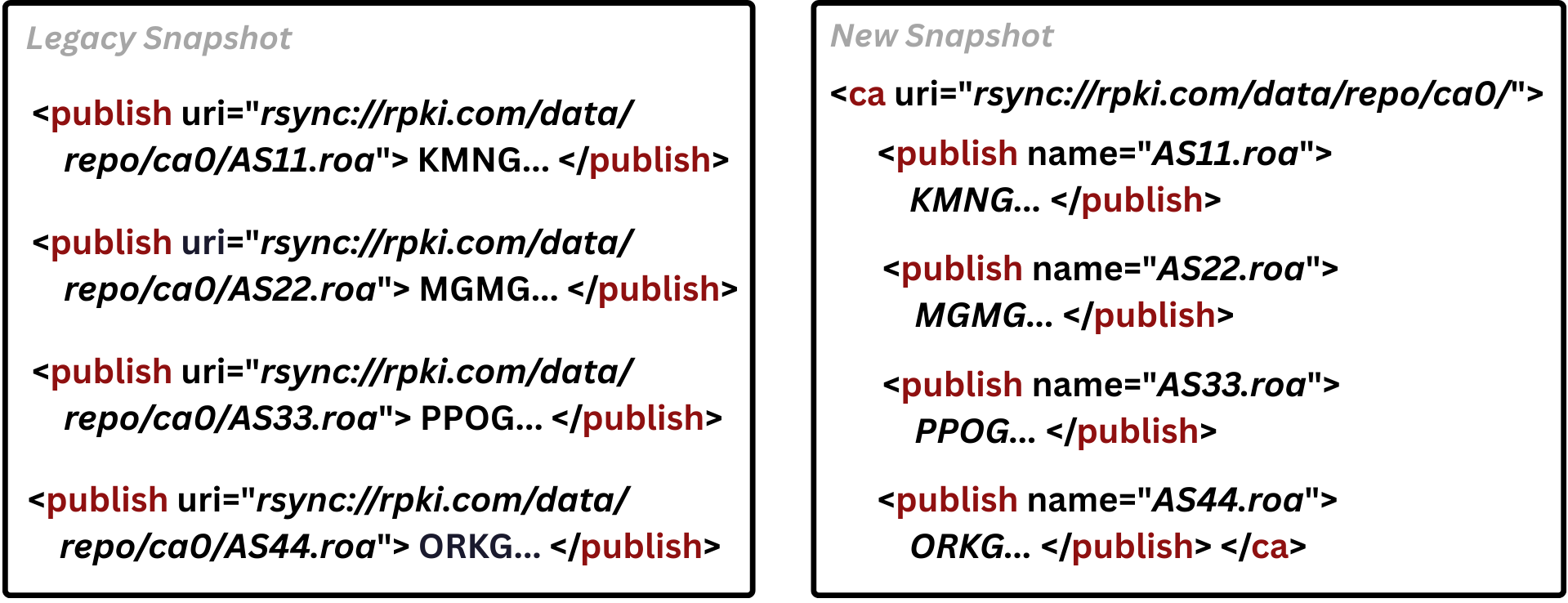}
    \vspace{-10pt}
    \caption{Comparison Snapshot structure}
    \label{fig:snap_compare}
    \vspace{-10pt}
\end{figure}

To overcome the full URI overhead, our improved snapshot template groups all objects of the same CA into one XML element that contains the repository URI in its header. Individual objects only include their respective name, which can be combined with the repository URI from the header to construct the full object URI. This optimization results in significantly smaller overhead (69\% smaller overhead for the improved ROAs) than storing the full URI for each object. The new design additionally increases readability, as all objects by the same CA are stored within the same (sub-)element.
The current / new snapshot design is illustrated in Figure~\ref{fig:snap_compare}.

\textbf{Protobuf over XML.}
Even with the improvements, XML elements still result in overhead. Human readability of XML structure elements, like $<$snapshot$>$, take up significant space, and the requirements for XML parsing and validation create substantial processing effort in the RPs (11.5\% of total processing time in Routinator).
Using a different encoding format than XML improves size and processing speed of RRDP files.  Looking at previous research \cite{gligoric2011performance, sumaray2012comparison}, we find protobuf as the most promising solution to improve RRDP files, as it enables smaller file sizes and more efficient parsing than XML. 
Using our above derived optimizations, we develop protobuf formats for all RRDP files, derived from the existing improved XML structure. We provide protobuf definitions for all RRDP files in Appendix~\ref{app:protorrdp}. {\updated Using protobuf additionally removes susceptibility of RPs to XML-based attacks, like XML-bombs \cite{spath2016sok}, but makes them potentially vulnerable to attacks on protobuf parsing. Keeping software up-to-date to prevent attacks on parsing thus remains important also in iRPKI.}

\subsection{Improved ASN.1 Object Encoding}
RPKI currently uses ASN.1 to encode objects. Our experiments show that encoding RPKI objects with protobuf can improve efficiency of object parsing. Due to the similar structure definitions of ASN.1 and protobuf syntax, adaptions to objects are relatively straight forward. All ASN.1 fields can be translated to protobuf equivalents. 

The downside of protobuf compared to ASN.1 is reduced flexibility. While ASN.1 generally allows for extensions, a wide range of types and flexible encoding, protobuf requires strict object definitions, has limited types and only supports one specific encoding. 
However, this strictness of protobuf is not a limitation in the context of RPKI. The RPKI RFCs specifically express that RPKI is designed to be strict and tailored to its very specific use-case \cite{rfc6487}. The RFCs only allow specific types, one encoding, and disallow any extensions or optional fields. This ensures deterministic structure of RPKI objects, preventing real-world object structure to diverge overtime. The strictness also omits the benefits of using a technology like ASN.1, that supports many other use-cases and provides templates for other object types, which are neither necessary nor desired by the RFCs in the context of RPKI. 
We provide protobuf definitions for improved ROAs and improved manifests in Appendix~\ref{app:protoroa} and \ref{app:protomft}.

\subsection{Backwards Compatibility}
The proposed RPKI design changes are not backwards compatible with current RPKI implementations, and parsing/validation will fail if new object structures are used where the RP expects the legacy once. We thus propose to implement the updated design in objects with new extensions.

All updates to the manifest file, i.e., removing EE-certificate, integrating the CRL and encoding with protobuf are incorporated into a new .imft file. Legacy RPs encountering a .imft ignore the file with the unknown extension, while updated RPs can processes the new manifest instead of the .mft file. 
Similarly, all other objects use an updated extension to indicate new object template, like .iroa. For RRDP files, which usually have a .xml extension, we propose a .bin extension. 

Simply including these objects into existing repository structure does not provide a benefit, as improved RPs still need to download and validate legacy objects to find the improved objects, and legacy RPs need to additionally download all new objects without being able to process them. We thus propose new objects are added into a parallel version of the repository. {\updated An RP supporting iRPKI requests a Notification.bin file at the same folder URI as the Notification.xml, e.g., \textit{../rrdp/notification.bin} instead of \textit{../rrdp/notification.xml}. This updated RRDP file and its corresponding Snapshot / Deltas contain all updated objects. If a Notification.bin is not found (404), the RP falls back to regular RPKI to download and process the legacy Notification.xml file.} This design only results in minimal overhead for the RPs, which generally do not get stalled from the failed request. Experiments shows that Fort is the only RP that will stall for 4 seconds if the server returns a 404 error, compared to milliseconds in Routinator and rpki-client. Implementing iRPKI support into Fort must ensure that a poll for Notification.bin does not result in stalling and the RP should instead immediately move to downloading the legacy Notification.xml. This is possible by setting a smaller request timeout for requests to the Notification.bin.

Using this parallel design ensures that all legacy RPs are not impacted by new RPKI objects, while updated RPs can utilize the performance benefits of iRPKI. Since repositories are separate and object extensions are unique, introducing iRPKI to live RPKI does not break any existing deployments, and RPs / repositories can incrementally move to iRPKI. RP operators benefit from deploying the new design, as bandwidth and processing requirements are reduced for each repository that provides iRPKI content. Likewise, repository operators benefit from lower bandwidth requirements for each RP fetching only iRPKI content. Using iRPKI thus provides incentives for both RP and repository operators, facilitating deployment.

\section{Evaluations}\label{sec:evaluation}
{\updated We implement all improvements derived in Section~\ref{sec:improvements} into Routinator 0.14.1 to illustrate the performance benefits. All iRPKI changes are backwards-compatible: The improved version of Routinator remains fully functional for current RPKI, and additionally supports new object types and validation concepts.} In total, adding support for all iRPKI functionality and objects required adding 1836 lines of code to Routinator. To create iRPKI objects and construct them into a repository, we use the open-source RPKI tool CURE \cite{mirdita2023cure}.
We adapt CURE to support legacy and improved versions of each object, which can be selected through CLI arguments.  

\subsection{Setup}
As RPKI is deployed in production systems, testing on real-world RPKI is infeasible. We instead construct a local isolated environment to run all our evaluations that is not connected to real-world RPKI. All tests use Ubuntu 24.04LTS with 32 GB RAM, and an 8-core current generation i7 processor. Nginx is used to serve the RRDP files. Following Section~\ref{sec:improvements}, we do not limit the bandwidth of the nginx webserver, as real-world available bandwidth can vary between servers, and size comparison between RPKI and iRPKI provides a clearer picture on bandwidth requirements.

\subsection{Repository Structure}
Evaluating the impact of individual improvements is not trivial, as benefits depend on the structure of repositories: A CA with many ROAs will benefit more from ROA improvements than from manifest improvements, while a CA without any ROAs will not benefit from ROA improvements at all. Experiments show that real-world CAs are diverse, with a minimal ROA amount of 0, and a maximum of 48886. 
To illustrate benefits for CAs with many ROAs vs. CAs with few ROA, we run each evaluation test-case with two repository structures: To test processing of payloads, like ROAs, we create a single CA setup that includes 10,000 ROA files. We chose 10,000 ROAs as we did not find any changes in our comparative results with more than 10,000 ROAs. 

For our second setup, we create 10,000 CAs with one CA-certificate, manifest, CRL, and ROA each, allowing us to evaluate improvement impact for CAs with few ROAs.

A further challenge in evaluation is the overlap of results. When testing all combined improvements, identifying the contributions of individual improvements is not possible. We thus run all test-cases with only the individual improvement. For example, when testing impact of protobuf encoding on RRDP download / processing times, we keep the legacy structure and validation of all other objects. 
We evaluate the combined effect of all improvements in test-case \textit{full}. 

\textbf{Real-world evaluation.}
Running our evaluations in a isolated environment is essential to provide clear and repeatable comparisons of the different optimizations, and to not cause issues in real-world RPKI operation. To still evaluate real-world impact of our improvements, we additionally conduct experiments with an emulated real-world RPKI structure. For this, we measure the amount and structure of all CAs that are currently part of the RPKI, and re-create their structure in our local environment, including 45k CA certificates, manifest, CRLs, and 290k ROAs. To run the experiments locally, we create our own TAL file pointing to our own local domain. Since we do not have access to the root private keys of RIRs, we use our own private root keys. This requires us to re-sign all RPKI objects. Further, we restore the validity of all objects for each test-case. This is necessary, as objects might otherwise expire during the experiments, potentially skewing results.

Our experiments fully reproduce the CA structure, including child-parent relationships and content in repositories. For example, the root ARIN CA includes 48,886 ROAs in our real-world measurement and in our local evaluation setup.

\subsection{Evaluation Metrics.}
Our evaluations include multiple metrics to rate the impact of our improvements.

\textbf{Total processing time.} 
We time the execution of the RP to evaluate how our improvements impact the total processing time. Performance benefits can stem from multiple parts of the processing, including smaller download times and parsing due to smaller objects or improved structure, faster validation due to lower cryptographic overhead, and lower cache storage times. We measure execution time with the Linux tool \textit{time}. 

\textbf{Bandwidth.}
To measure the bandwidth requirement of setups, we evaluate the total size of all RRDP files that the RPs need to download during a fetch. These files additionally provide insights into the memory requirements for repositories.

\textbf{RP cache.} 
We evaluate the memory footprint of the RP by checking the size of the Routinator cache folder after each test, providing insights into size impact on RP storage.

\subsection{Results}
\begin{figure}[t!]
    \centering
    \includegraphics[width=0.9\linewidth]{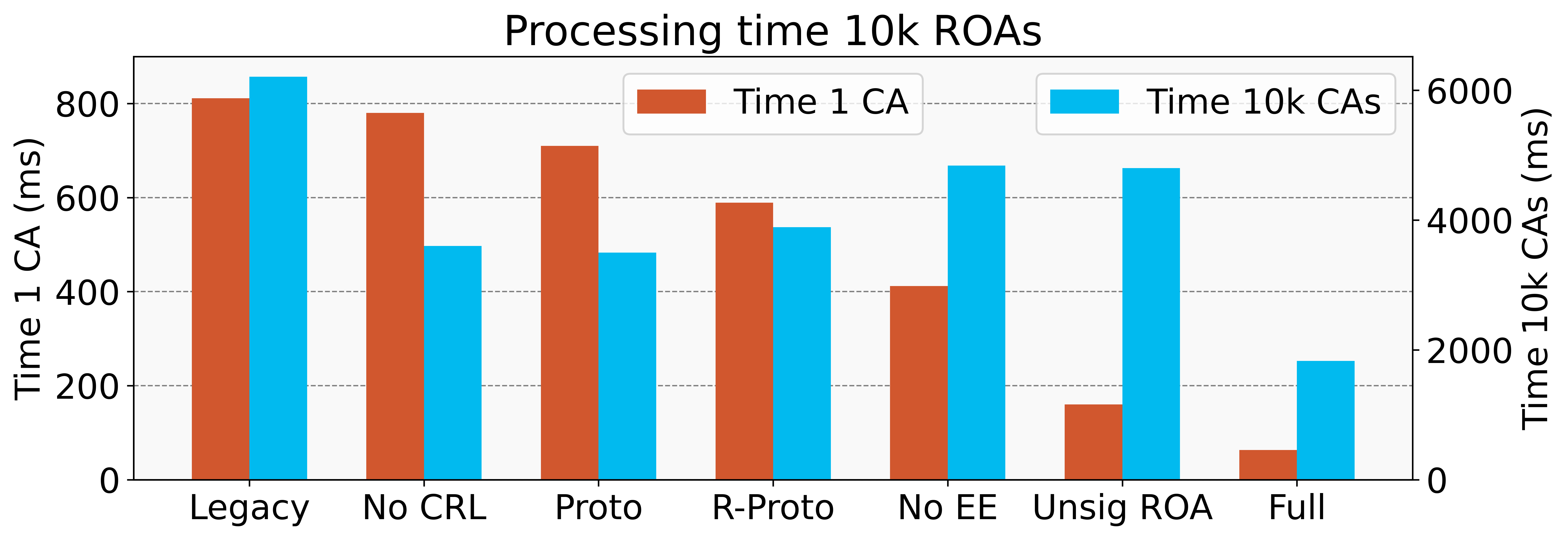}
    \vspace{-5pt}
    \caption{Processing time single 10k ROAs.}
    \label{fig:impr_sc}
    \vspace{-8pt}
\end{figure}

The results of our evaluations are shown in Figure~\ref{fig:impr_sc}. 

\textbf{No CRL.}
Removing the CRL has a significant impact on the test-case with many CAs, as it reduces the parsing effort, and the amount of required signature validations. The impact in the single CA case is limited, as it only includes one CRL, while a much more significant improvement can be observed for the 10k CA case. Removing the CRL reduces RRDP file size by 0.01\% in the 1 CA test, and 22\% for 10k CAs. 

\textbf{Proto over ASN.1.} 
Moving ROA / MFT object definitions from ASN.1 to protobuf has a measurable effect on total processing times, reducing the overall file-size by 15\% and speeding up object parsing by about 10\%. As protobuf encoding only impact parsing and file-size, not validation effort, we do not see any improvements in the validation time of the RP.

\textbf{Proto over XML.}
We see a more significant impact in moving from XML to protobuf for RRDP file encodings, with an improvement of 27\% in file size and an improvement of 31\% in parsing time. The encoding reduces redundancies to achieve both smaller files and faster processing.

\textbf{No EE-Certificates.}
Removing the EE-certificates from ROAs and manifests significantly reduces processing time in the single CA case by 52\%. The benefit is smaller for for the 10k CA test, with an improvement of around 36\%. We also observe that this optimization reduces RRDP file size by about 63\% and Routinator cache size by about 60\%, illustrating the substantial overhead introduced by the EE-certificates.

\textbf{No ROA signatures.}
Additionally removing ROA signatures increases the impact of optimizations, reducing processing time by 82\% for the single CA case. The immense impact stems from the amount of signature validations, which is reduced from 20,004 for the legacy RPKI to 3 (1 CRL, 1 mft, 1 CA-cert). The RP now spends the majority of time on parsing and processing objects, over signature validations. For the 10,000 CA case, the speed-up is not as significant, with about 41\%, as the RP still has to processing mfts, CRLs and CA-certificates. The improvement leads to a 91\% size reduction for single CA, and 51\% for the 10k CAs test. 

\textbf{Combined improvements.}
Combining all optimizations results in a substantial speed-up. For the single CA case, the processing time is reduced from 0.841s to 0.058s (93\%). Investigating the interaction between different improvements, we find that the impact of improvements influences each other. For example, moving from XML to protobuf for RRDP achieved an improvement of 31\% in parsing time when applied to legacy RPKI. Repeating this test for iRPKI, with all other improvements applied, we find that not using protobuf leads to a parsing time of 101.4ms, while applying protobuf brings down parsing time to 11.1ms, an improvement by 90\%. This larger impact results from the reduced file-size of RPKI objects: With smaller files in the Snapshot, the XML elements take up more overall space. Further, with smaller files, the validation of XML files takes up a larger percentage of total processing time, increasing the advantage of protobuf.

The combined improvements reduce the total RRDP file size for the single CA test from 23.6MB to 1.3 MB (95\%) and the Routinator cache size from 36.3 MB to 4.4 MB (88\%), illustrating the substantial benefits in file-size from iRPKI. 

For the 10,000 CA case, we observe smaller overall benefit, with a processing time reduction by about 70\%. The smaller improvement is expected, as the large amount of CAs still requires full CA certificates, limiting the overall improvement gains. Still, processing time is reduced by 3.7x.

\textbf{Bandwidth test.}
We additionally illustrate the benefits of iRPKI in a bandwidth constrained setup. This test is motivated by the observation that the target of RPKI is global deployment, but bandwidth can be limited in different settings, e.g., in less developed nations or low-resource configurations. For this test, we setup nginx with a bandwidth limitation of 3.23Mbit/s, the lowest global average Internet bandwidth (Syria)\footnote{\url{https://www.speedtest.net/global-index} (Accessed Mar 21 2025)}.

Our results show that legacy RPKI requires 7.9s for download and validation, while iRPKI takes 0.43s, an improvement by 95\%, illustrating that iRPKI substantially increases feasibility of running RPs in bandwidth constrained environments.

\begin{figure}[t!]
    \centering
    \includegraphics[width=0.9\linewidth]{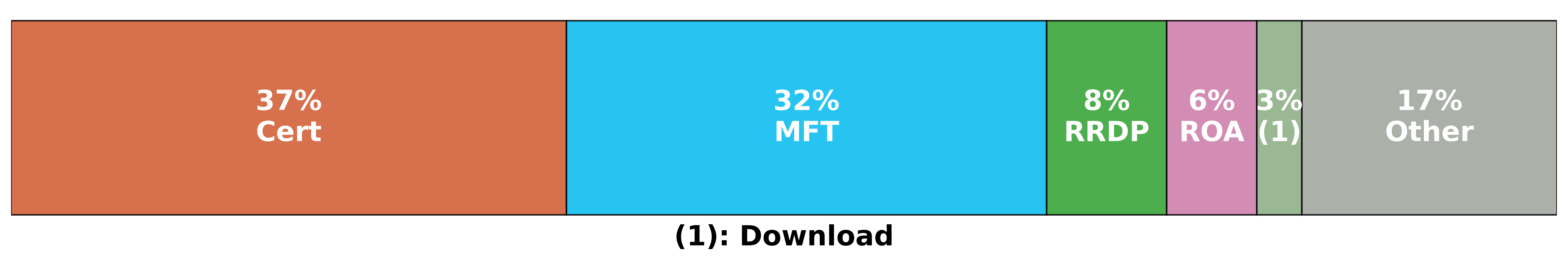}
    \vspace{-5pt}
    \caption{Flamegraph iRPKI.}
    \label{fig:flame_irpki}
    \vspace{-8pt}
\end{figure}

\textbf{Flamegraph iRPKI.}
We run a flamegraph evaluation using the same parameters as in Section~\ref{sec:background} on a full iRPKI setup, shown in Figure \ref{fig:flame_irpki}. In the optimized version, certificate processing takes up most processing time. This is expected, as certificates were intentionally not adapted in iRPKI to ensure compatibility with existing X.509 implementations, and ensure no potential issues with certificate path security are introduced. Manifest processing still requires a signature validation, leading to a remaining large fraction of processing. RRDP processing requires some parsing time, as the Snapshot remains relatively large, containing all repository objects. With the optimizations, processing ROAs is merely reduced to parsing the small objects and storing the content in the local state. In iRPKI, processing ROAs takes up minimal processing time, while it took up \~50\% of processing in legacy RPKI. 

The flamegraph illustrates that iRPKI reduces processing time to the required minimum, with most processing in security critical signature validations on CA-certs and manifests.

\subsection{Real-World Evaluations}
\begin{figure}
    \centering
    \includegraphics[width=0.7\linewidth]{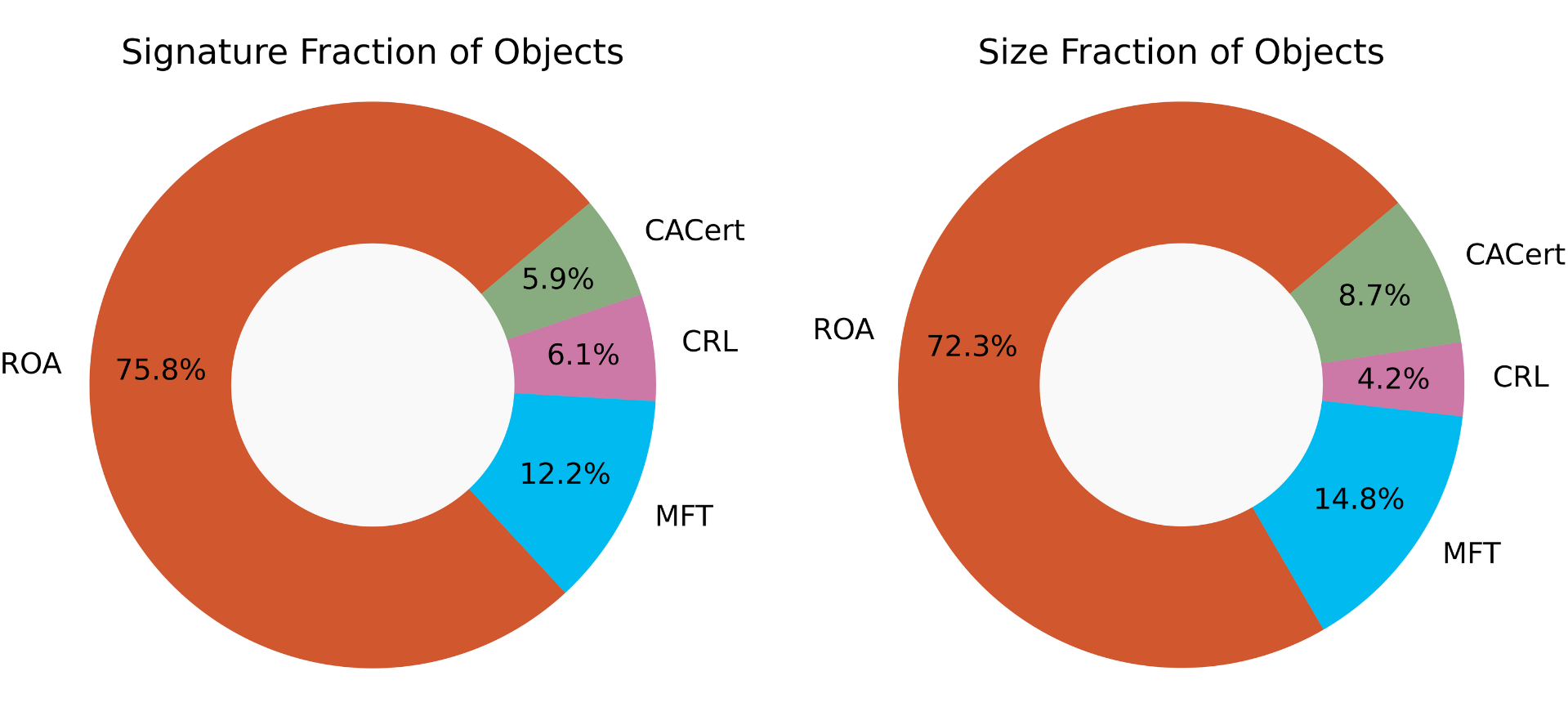}
    \vspace{-8pt}
    \caption{Fraction of different RPKI object types.}
    \label{fig:size_frac}
    \vspace{-10pt}
\end{figure}

We additionally evaluate iRPKI on a setup derived from real-world RPKI structure. On the day of our measurement, Jan 28 2025, the global RPKI contains 427,937 objects. 
Figure \ref{fig:size_frac} illustrates how each type impacts the total (binary) size and total signature validations of current RPKI objects. The figures show that ROAs make up the largest part of all RPKI data and of all required signature validations. {\updated A comparison to iRPKI can be found in Appendix~\ref{app:fraction}.}

To evaluate how iRPKI would impact real-world objects, we process the current RPKI state to convert all objects to their iRPKI counterparts. All ROAs are stripped of their EE-certificate and signature, and serial-number and expiration time are added to the content. The EE-certificate is removed from the manifest, and the content of the respective CRL is added to the manifest. Manifests and ROAs are encoded with protobuf. Notification and Snapshots are adapted to the new improved structure and also encoded with protobuf. 

We host all files in a local isolated RPKI tree to ensure our measurements do not impact real-world RPKI. Finally, we setup an nginx to serve the RRDP files. To ensure our setup emulates the real-world RPKI accurately, we measure the average data rate of RPKI downloads across all global repositories, which is around 40 Mbit/s. We then limit the data-rate of nginx to 40 Mbit/s. We run Routinator three times and average results to ensure reproducibility.

\textbf{Results.} 
In our first test, we run Routinator against the emulation of the current, unaltered RPKI state. Routinator finishes the evaluations in 328s. The faster fetch compared to real-world RPKI despite similar bandwidth (328s vs. 402s) is due to connection errors and timeouts during real-world downloads, which do not occur in the local isolated setup. 

We then run a test with the full current RPKI state converted to iRPKI. With the improved version, Routinator finishes in 57s, 5.8x faster than current RPKI. The total RRDP snapshot size is reduced from 1.1 GB to 160 MB, a 6.9x improvement. We additionally include delta bandwidth evaluation in Appendix~\ref{app:delta_fetches}.
Since iRPKI does not make changes to object validity, both states lead to the same VRPs output. 
The real-world evaluations illustrate the substantial benefit of iRPKI when the full current RPKI state is converted to iRPKI. 

\section{Path Forward: From Design to Deployment}\label{sec:discussion}
Our roadmap for iRPKI requires:

\textbf{RFCs.} Implementation of iRPKI requires changes to RFC specification. We list all RFCs that are affected by our changes in Appendix~\ref{app:rfclist}. The most suitable group within the IETF to work on iRPKI changes is the SIDROPS working group, responsible for standardization of new RPKI RFCs.
The proposed changes break the (unimplemented) RPKI use-case of out-of-band distribution of objects, e.g., through BGP messages. While no such use-case was developed, keeping the ability to support other forms of distribution in the future is worthwhile. Standardizing the proposed changes within new independent RFCs ensures existing templates remain valid and usable once fitting use-cases arise. 

\textbf{Timeline.} Substantial changes to Internet protocols are traditionally slow, and can take years to incorporate into the specification. For example, incorporating ASPA into RPKI has been ongoing for seven years. To facilitate a swift inclusion of our improvements into the specification, we propose starting with easier-to-implement changes first. 
For example, discussions on a new RRDP version are already ongoing \cite{mailingListRRDP}, and could incorporate the changes to the file structure and the encoding proposed in this work. Changes to the fundamental structure and encoding of RPKI objects are essential but require modifying the current object structure and validation specification, and will therefore incur a longer process. 

{\updated
\textbf{Hybrid deployment.}
During the transition from  RPKI to iRPKI, both technologies will co-exist. This creates additional load for repositories supporting both versions:  Repositories must retain all existing objects for legacy RPs, and offer iRPKI objects for updated RPs; a storage overhead of about 14.5\%. Supporting iRPKI allows the repository to serve significantly smaller snapshots (6.9x on average) to any RP supporting iRPKI. Following Section~\ref{sec:problems}, bandwidth is a larger operational concern for repositories than storage and iRPKI thus provides operational benefits even during hybrid deployment.
For RPs, the only overhead is potential failed requests to Notification.bin files for all repositories not supporting iRPKI, which currently amounts to a maximum of 91 requests. RPs supporting iRPKI get the bandwidth, computation and storage improvements for any iRPKI-supporting repository.}

\section{Conclusion}\label{sec:conclusion}
To deliver on its security promises, RPKI requires wide deployment, and support for many use-cases, like ASPAs for leak prevention or BGPsec certificates for BGP path validation. In its current design, RPKI lacks the scalability to support the required amount of objects to achieve full deployment for all such mechanisms. In addition, the complexity of RPKI architecture lead to software vulnerabilities. With increasing RPKI deployment, the margins of resilience to networking problems or attacks will substantially decrease.

To overcome these limitations, fundamental changes to the architecture of RPKI are necessary. In this work, we identify substantial inefficiencies in the design of RPKI and develop specification-level improvements to overcome them. We demonstrate a 6.9x size improvement and 5.8x faster processing when the current real-world RPKI state is converted to iRPKI. 
iRPKI is backwards compatible with RPKI deployments, and offers benefits even to early adopters. We expect that these properties will contribute to speeding up the required adaptions to the specification and the standardization process. With iRPKI, the RPKI architectures becomes scalable, enabling support for new and future use-cases, operation in constrained environments, and promises to remain functional in full global deployment.
To facilitate deployment and changes to specification and enable reproduction of our study, we open-source exemplary iRPKI objects, the adapted Routinator code, and the repository tool\footnote{\href{https://www.dropbox.com/scl/fi/pvs6odh2ml25azifd6vg9/submission_ndss.zip?rlkey=eadvjuoxo7tar1q82h08cb4fu&st=elqnsfix&dl=0}{www.dropbox.com/scl/fi/pvs6odh2ml25azifd6vg9/submission\_ndss.zip}}.

\bibliographystyle{plain}
\bibliography{NetSec}

\begin{thebibliography}{10}

\bibitem{rfc6486}
Rob Austein, Geoff Huston, Stephen Kent, and Matt Lepinski.
\newblock Manifests for the resource public key infrastructure (rpki).
\newblock Technical report, 2022.

\bibitem{bishop2008ethical}
Matt Bishop and David~V. Bailey.
\newblock Ethical issues in computer security research: What practitioners say.
\newblock In {\em Proceedings of the 4th Workshop on Ethics in Computer Security Research (WECSR)}. Springer, 2008.

\bibitem{bcpserver}
T~Bruijnzeels, Ties De~Kock, F~Hill, T~Harrison, and J~Snijders.
\newblock Rpki publication server best current practices.
\newblock 2025.

\bibitem{rfc8182}
Tim Bruijnzeels, Oleg Muravskiy, Bryan Weber, and Rob Austein.
\newblock The rpki repository delta protocol (rrdp), 2017.

\bibitem{klayswap}
Catalin Cimpanu.
\newblock { KlaySwap crypto users lose funds after BGP hijack}.
\newblock \\\url{https://therecord.media/klayswap-crypto-users-lose-funds-after-bgp-hijack}, 2022.
\newblock Accessed 09/09/2024.

\bibitem{rfc5280}
David Cooper, Stefan Santesson, Stephen Farrell, Sharon Boeyen, Russell Housley, and William Polk.
\newblock Internet x. 509 public key infrastructure certificate and certificate revocation list (crl) profile.
\newblock Technical report, 2008.

\bibitem{mailingListRRDP}
Ties de~Kock.
\newblock Ietf sidrops publication point rp synchronization in bandwidth constrained environments (note for rrdp v2).
\newblock \\\url{https://mailarchive.ietf.org/arch/msg/sidrops/zby0IyetZPyeL6606hUVN1dViKg}, 2023.
\newblock Accessed Jul 7 2025.

\bibitem{gilbert2003security}
Henri Gilbert and Helena Handschuh.
\newblock Security analysis of sha-256 and sisters.
\newblock In {\em International workshop on selected areas in cryptography}, pages 175--193. Springer, 2003.

\bibitem{gligoric2011performance}
Nenad Gligori{\'c}, Igor Dejanovi{\'c}, and Sr{\dj}an Kr{\v{c}}o.
\newblock Performance evaluation of compact binary xml representation for constrained devices.
\newblock In {\em 2011 international conference on distributed computing in sensor systems and workshops (DCOSS)}. IEEE, 2011.

\bibitem{cloudfarehijack}
Bryton Herdes, Mingwei Zhang, and Tanner Ryan.
\newblock { Cloudflare 1.1.1.1 incident on June 27, 2024}.
\newblock \\\url{https://blog.cloudflare.com/cloudflare-1111-incident-on-june-27-2024/}, 2024.
\newblock Accessed 09/09/2024.

\bibitem{hlavacek2023beyond}
Tomas Hlavacek, Philipp Jeitner, Donika Mirdita, Haya Shulman, and Michael Waidner.
\newblock Beyond limits: How to disable validators in secure networks.
\newblock In {\em Proceedings of the ACM SIGCOMM 2023 Conference}, 2023.

\bibitem{hlavacek2023keep}
Tomas Hlavacek, Haya Shulman, Niklas Vogel, and Michael Waidner.
\newblock Keep your friends close, but your routeservers closer: Insights into $\{$RPKI$\}$ validation in the internet.
\newblock In {\em 32nd USENIX Security Symposium (USENIX Security 23)}, 2023.

\bibitem{rfc9364}
Paul~E Hoffman.
\newblock Dns security extensions (dnssec).
\newblock {\em RFC 9364}, 2023.

\bibitem{whitehouse2024routing}
The~White House.
\newblock {Roadmap to Enhancing Internet Routing Security}, September 2024.

\bibitem{rfc5652}
Russell Housley.
\newblock Rfc 5652: Cryptographic message syntax (cms), 2009.

\bibitem{rfc6485}
G~Huston.
\newblock The profile for algorithms and key sizes for use in the resource public key infrastructure (rpki), 2012.

\bibitem{apnicrov}
Geoff Huston.
\newblock {APNIC ROV World Map}, 2025.
\newblock Accessed Jul 3 2025.

\bibitem{rfc6487}
Geoff Huston, George Michaelson, and Robert Loomans.
\newblock A profile for x. 509 pkix resource certificates, 2012.

\bibitem{rfc6488}
Matt Lepinski, Andrew Chi, and Stephen Kent.
\newblock Signed object template for the resource public key infrastructure (rpki).
\newblock Technical report, 2012.

\bibitem{rfc6480}
Matt Lepinski and Stephen Kent.
\newblock An infrastructure to support secure internet routing.
\newblock Technical report, 2012.

\bibitem{li2023rovista}
Weitong Li, Zhexiao Lin, Md~Ishtiaq Ashiq, Emile Aben, Romain Fontugne, Amreesh Phokeer, and Taejoong Chung.
\newblock Rovista: Measuring and analyzing the route origin validation (rov) in rpki.
\newblock In {\em Proceedings of the 2023 ACM on Internet Measurement Conference}, 2023.

\bibitem{liu2015end}
Yabing Liu, Will Tome, Liang Zhang, David Choffnes, Dave Levin, Bruce Maggs, Alan Mislove, Aaron Schulman, and Christo Wilson.
\newblock An end-to-end measurement of certificate revocation in the web's pki.
\newblock In {\em Proceedings of the 2015 Internet Measurement Conference}, 2015.

\bibitem{macnish2020ethics}
Kevin Macnish and Jeroen Van~der Ham.
\newblock Ethics in cybersecurity research and practice.
\newblock {\em Technology in society}, 63, 2020.

\bibitem{bgp:attacks}
Carolyn~Duffy Marsan.
\newblock Six worst internet routing attacks, 2009.

\bibitem{mirdita2023cure}
Donika Mirdita, Haya Schulmann, Niklas Vogel, and Michael Waidner.
\newblock The cure to vulnerabilities in rpki validation.
\newblock 2024.

\bibitem{mirdita2024sok}
Donika Mirdita, Haya Schulmann, and Michael Waidner.
\newblock {SoK: An Introspective Analysis of RPKI Security}.
\newblock {\em 34st USENIX Security Symposium (USENIX Security 25)}, 2025.

\bibitem{mirdita2022poster}
Donika Mirdita, Haya Shulman, and Michael Waidner.
\newblock Poster: Rpki kill switch.
\newblock In {\em Proceedings of the 2022 ACM SIGSAC Conference on Computer and Communications Security}, 2022.

\bibitem{rfc1035}
Paul~V Mockapetris.
\newblock Rfc1035: Domain names-implementation and specification, 1987.

\bibitem{morillo2021rov}
Reynaldo Morillo, Justin Furuness, Cameron Morris, James Breslin, Amir Herzberg, and Bing Wang.
\newblock Rov++: Improved deployable defense against bgp hijacking.
\newblock In {\em NDSS}, 2021.

\bibitem{nistmonitor}
NIST.
\newblock Nist rpki monitor.
\newblock 2025.
\newblock Accessed Jul 7 2025.

\bibitem{rovnlnetlabs}
NLNetLabs.
\newblock Nlnetlabs rpki statistic.
\newblock \\\url{https://rov-measurements.nlnetlabs.net/stats/}, 2025.
\newblock Accessed Jul 7 2025.

\bibitem{rfc8446}
Eric Rescorla.
\newblock The transport layer security (tls) protocol version 1.3.
\newblock Technical report, 2018.

\bibitem{singh2019taxonomy}
Ankit Singh, Aditi Sharma, Nikhil Sharma, Ila Kaushik, and Bharat Bhushan.
\newblock Taxonomy of attacks on web based applications.
\newblock In {\em 2019 2nd International Conference on Intelligent Computing, Instrumentation and Control Technologies (ICICICT)}, volume~1. IEEE, 2019.

\bibitem{mailingListReview}
Job Snijders.
\newblock Ietf sidrops rpki's 2024 year in review.
\newblock \\\url{https://mailarchive.ietf.org/arch/msg/sidrops/wI_PqEMsScRh1-jYl8XYPDI-3qE/}, 2025.
\newblock Accessed Jul 7 2025.

\bibitem{spath2016sok}
Christopher Sp{\"a}th, Christian Mainka, Vladislav Mladenov, and J{\"o}rg Schwenk.
\newblock $\{$SoK$\}$:$\{$XML$\}$ parser vulnerabilities.
\newblock In {\em 10th USENIX workshop on offensive technologies (WOOT 16)}, 2016.

\bibitem{mailingListLatency}
Yingying Su.
\newblock Ietf sidrops draft-li-sidrops-rpki-repository-problem-statement-00-scalability issues.
\newblock \\\url{https://mailarchive.ietf.org/arch/msg/sidrops/86t7KlugB8v08dX-j9hHDxE8Ubs}, 2024.
\newblock Accessed Jul 7 2025.

\bibitem{su2024drr}
Yingying Su, Dan Li, Li~Chen, Qi~Li, and Sitong Ling.
\newblock drr: A decentralized, scalable, and auditable architecture for rpki repository.
\newblock In {\em NDSS}, 2024.

\bibitem{sumaray2012comparison}
Audie Sumaray and S~Kami Makki.
\newblock A comparison of data serialization formats for optimal efficiency on a mobile platform.
\newblock In {\em Proceedings of the 6th international conference on ubiquitous information management and communication}, 2012.

\bibitem{DBLP:journals/cacm/SunABVRCM21}
Yixin Sun, Maria Apostolaki, Henry Birge{-}Lee, Laurent Vanbever, Jennifer Rexford, Mung Chiang, and Prateek Mittal.
\newblock Securing internet applications from routing attacks.
\newblock {\em Commun. {ACM}}, 64(6):86--96, 2021.

\bibitem{van2022rpkiller}
Koen Van~Hove, Jeroen van~der Ham, and Roland van Rijswijk-Deij.
\newblock Rpkiller: Threat analysis from an {RPKI} relying party perspective.
\newblock {\em Digital Threats: Research and Practice Vol. 4}, 2023.

\end{thebibliography}

\section{Ethical Considerations}
We ensure our work is ethical by conducting all tests in isolated environments. We do not test our improved design on the production RPKI infrastructure but reproduce the state in our lab. In our work, we following best practices for research in network and software security \cite{macnish2020ethics, bishop2008ethical}.
\begin{appendix}
    \subsection{Requests RPs}\label{app:poll_times}
\begin{figure}[h]
    \centering
    \includegraphics[width=0.95\columnwidth]{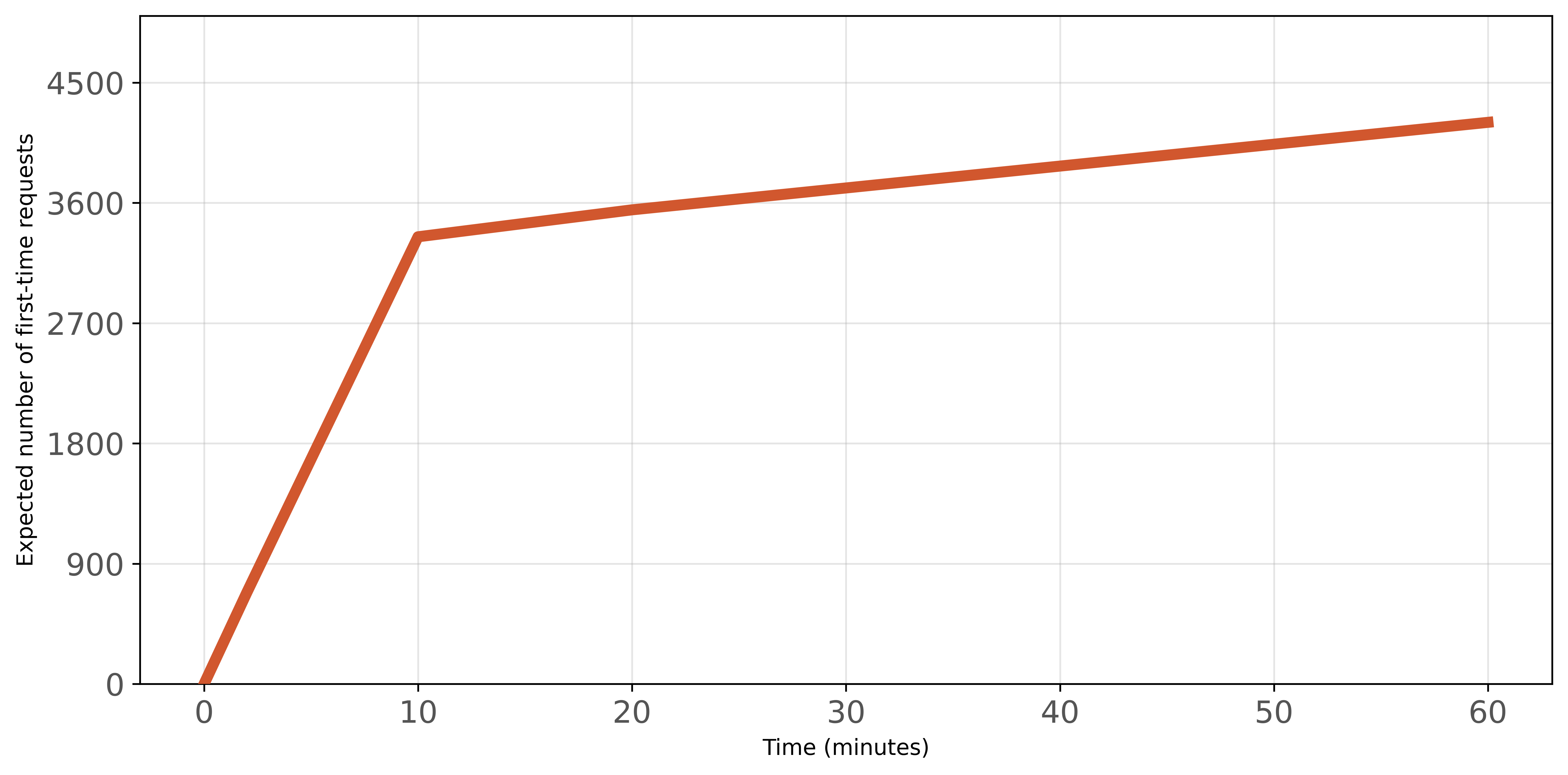}
    \vspace{-5pt}
    \caption{Average amount requests.}
    \vspace{-15pt}
    \label{app:fig:poll_times}
\end{figure}

{\updated Using the default RP fetch intervals, we can approximate the time distribution of the amount of RPs that request the repository for the first time after a session reset, illustrated in Figure~\ref{app:fig:poll_times}.
Within 10min, the average repository will have to serve all Routinator clients, 1/6 of all Fort and rpki-client instances and 58 other clients (e.g. OctoRPKI). On average, the snapshot has 15 MB (120Mbit). This yields (3125 + 757/6 + 255/6 + 58) * 120Mbit / 600s = 670.3Mbit/s.}
 
This data-rate is a minimal estimate, not accounting for HTTPS overhead, and retransmissions caused by congestion or connection errors.  
The repository also needs to account for RP fallback, which in a worse case can double the amount of data downloaded by one RP. Fallbacks occur if the connection or snapshot processing fail, prompting the RP to fall back to downloading the entire repository content over rsync. 
If bandwidth is insufficient in such cases, the repository can experience a negative spiral, as discussed in an IETF BCP and on the IETF mailing list\footnote{\href{https://datatracker.ietf.org/doc/draft-ietf-sidrops-publication-server-bcp/}{Link BCP} and \href{https://mailarchive.ietf.org/arch/msg/sidrops/wI_PqEMsScRh1-jYl8XYPDI-3qE/}{Link IETF}}. 
In such a negative spiral, a congested connection leads more and more requesting RPs to initiate retransmissions and fallback, continuously worsening the congestion. 
Repositories are thus advised to ensure sufficient bandwidth overhead to prevent such congestion spirals to occur.

\subsection{Fetch Thresholds}\label{app:fetch_issues}
All RPKI objects have expiration dates to prevent replay attacks with old files and ensure RPs have up-to-date data. Longer fetch times on RPs become problematic if they delay the update so much that objects expire before the run is finished. To test when this occurs, we download the current RPKI state and check the remaining time until expiration of the objects. We find that manifests and CRLs generally have the smallest expiration times, which is sensible, as they must always contain the current state of the repository and thus require the most frequent updates. If CRL or manifest expire, the CA becomes invalid. 
The median remaining validity of manifests and CRLs is 1303 minutes (21.7h), with a minimal expiration time of 135min. If an RP takes 135min to complete the fetch, the expiration check after the download will invalidate the object and exclude the entire CA from the VRPs. 
Thus, to ensure no validly signed objects are exclude, fetches must be finished within 135min in current RPKI. In practice, RPs can include additional timeouts that shorten this interval. For example, rpki-client terminates an update after 60min. If a fetch takes longer than 60min, data becomes unavailable. {\updated While current best-case fetch durations are still far from the 60min, previous work has shown configuration errors can extend fetch duration significantly \cite{mirdita2024sok}. Attacks on repositories can slow down download speed, further extending fetches, and the growth of RPKI will, inadvertently also extend fetch durations. Longer fetches lower the margin of error, and can thus lead to data-loss, as discussed in \ref{app:long_fetches}. 

Additionally to long fetches, overloaded repositories can additionally lead to data-loss. Monitoring availability of RPKI repositories from a well-connected network, we find 20,000 VRPs missing from Fort (measured on June 29 2025) since it enforces a minimal data-rate of 10kbit/s. This data-rate is not continuously provided by LACNIC, likely due to overload or active attacks, leading to unavailability of LACNIC data in Fort. With more RPKI deployment and correspondingly, more load on repositories, this problem will become worse.}

\subsection{Long Fetches}\label{app:long_fetches}
Long fetches decrease the resilience and flexibility of RPKI, as it lowers the margins for error and tolerance for unforeseen conditions. Even a repository that provides sufficient bandwidth in benign network conditions may face issues if congestion due to benign traffic or targeted attacks occurs, as the data in the repository can become unavailable\footnote{\href{https://datatracker.ietf.org/doc/draft-ietf-sidrops-publication-server-bcp/}{Link BCP}}. 
Due to these resource requirements, RPs can only be installed in well-connected, strong computation environments, limiting the flexibility for network operators how to deploy RPs in their system. With shorter fetch intervals and computation requirements, flexibility for operators increases.

Fetch times also impact operational considerations in RIRs. Since RPKI operation can be heavily impacted by misconfigured repositories, the RIRs need emergency processes in case misbehaving repository operators strain the system so much that it causes issues to RPs, requiring technical or legal intervention to prevent failures\footnote{\href{https://mailarchive.ietf.org/arch/msg/sidrops/GyFG6enaDn0zriThDZS9YDNbuos/}{Link IETF}}. With longer average fetch times, the margin after which the RIR have to activate emergency processes gets smaller. 

Further, increased fetch times increase the propagation delay between publication of a new RPKI object and the decision process of routers. For example, if a faulty ROA is issued that leads to traffic loss, correcting this mistake by revoking the ROA takes longer with extended RP fetch intervals. Such faulty ROAs are still prevalent in RPKI today\footnote{\url{https://rpki-monitor.antd.nist.gov/} (Accessed April 17 2025)}. 

\subsection{Post-Quantum Sizes}\label{app:postquant}
Currently, two finalists for post-quantum signatures are evaluated by NIST, ML-DSA and SLH-DSA\footnote{\href{https://www.nist.gov/news-events/news/2024/08/nist-releases-first-3-finalized-post-quantum-encryption-standards}{NIST finalists}}. If the smallest available option is chosen for RPKI, ML-DSA-44, public key size shifts from 256 byte to 1312 byte, and signature size increases from 256 byte to 2420 byte. Changing all current RPKI objects from RSA-2048 to ML-DSA-44 would increase total RPKI size from 1.2 GB to around 2.9 GB. While this is already a substantial increase, it assumes the best-case scenario regarding size. 
In a worst case, if SLH-DSA-SHA2-256f is chosen for RPKI signatures, signing all current RPKI objects with it would decrease public key size from 256 byte to 64 byte, but increase signature size from 256 byte to 49.86 kB, leading to a total RPKI size of 39.1 GB. While no final decision on the post-quantum algorithm has occurred to-date, the necessary eventual shift to stronger post-quantum secure algorithms will put a substantial strain on RPKI deployments. 

\subsection{Bandwidth Delta}\label{app:delta_fetches}
{\updated RPKI operation defaults to applying deltas during updates, if available. To evaluate iRPKI impact on deltas, we fetch all currently available delta files from live RPKI repositories. In total, we download 2863 deltas\footnote{Measured Jul 4 2025} and convert them to iRPKI. All objects are converted to their iRPKI equivalent and content is encoded with the delta protobuf template. 
In total, the size of deltas is reduced from 402 MB to 150 MB, a reduction by around 63\%. The size reduction is smaller than for snapshots due the different object distribution in deltas: While ROAs make up 72\% of global RPKI objects, they only make up 4\% of objects in deltas. In total, RPKI delta contains 53 certificates, 4308 ROAS, 51871 CRLs and 51879 manifests. Since iRPKI size reduction for ROAs is more significant than for other objects, the size reduction is smaller than for snapshots. 
The majority of delta updates only update validity of manifest and CRL, not issuing or modifying any ROAs.  Using iRPKI reduces the amount of necessary signature validations from 164304 to 51937, a reduction of 72\%. 
Using iRPKI thus has substantial benefits for bandwidth and validation effort, even if exclusively deltas are used.}

\subsection{Fraction iRPKI Objects}\label{app:fraction}
{\updated The size fractions in iRPKI can be seen in Figure~\ref{fig:frac_irpki}. Without ROA signatures, total signature validation efforts is rougly 50\% divided between CA certificates and manifest. The significantly smaller size of iRPKI objects is also evident in the figure: While ROAs made up 72\% of the total iRPKI size, the only take up 3.8\% in iRPKI. 

\begin{figure}
    \centering
    \includegraphics[width=1.0\columnwidth]{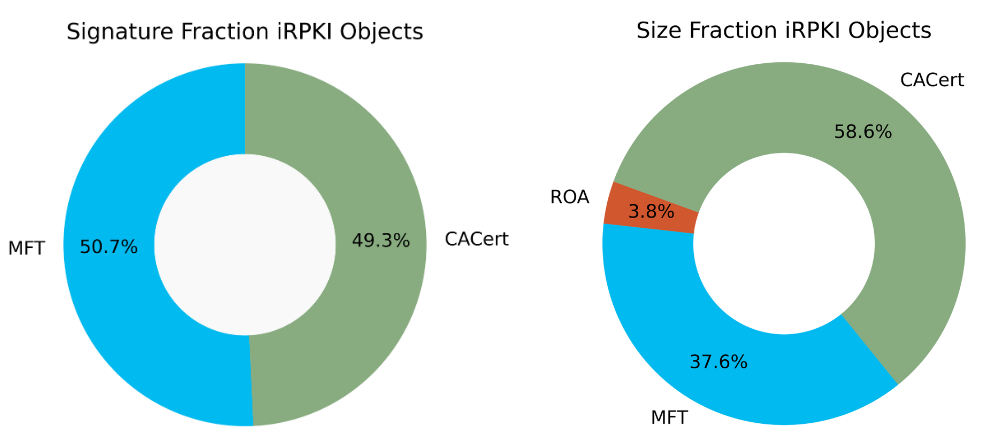}
    \caption{Fraction of signatures per object type (left) and fractions of total size per object type (right). }
    \label{fig:frac_irpki}
\end{figure}
}

\section{Object Templates ASN.1}\label{app:templates_asn1}
\subsection{Improved Manifest}\label{app:template_asn1_mft}
The improved manifest removes the EE-certificate and incorporates the CRL. Below, we provide the encapsulatedContent definition of the MFT.

\begin{lstlisting}[language=ASN1]
Manifest ::= SEQUENCE {
   version           [0] INTEGER DEFAULT 0,
   manifestNumber        INTEGER (0..MAX),
   thisUpdate            GeneralizedTime,
   nextUpdate            GeneralizedTime,
   fileHashAlg           OBJECT IDENTIFIER,
   fileList              SEQUENCE SIZE (0..MAX) OF FileAndHash
   revokedCertificates   SEQUENCE SIZE (0..MAX) OF revokedCertificate
}
\end{lstlisting}
\begin{lstlisting}
FileAndHash ::= SEQUENCE {
  file  IA5String,
  hash  BIT STRING
}
\end{lstlisting}

\begin{lstlisting}
revokedCertificate ::= SEQUENCE  {
    userCertificate         CertificateSerialNumber,
    revocationDate          Time,
}
\end{lstlisting}

\begin{lstlisting}
CertificateSerialNumber  ::=  INTEGER
\end{lstlisting}

\subsection{Improved ROA}\label{app:template_asn1_roa}
The improved ROA removes EE-certificate and signerInfos, gets rid of the signedData type and incorperates a ROA number and a validity period.

\begin{lstlisting}[language=ASN1]
RouteOriginAttestation ::= SEQUENCE {
    version [0]           INTEGER DEFAULT 0,
    asID                  ASID,
    roaNumber             CertificateSerialNumber
    thisUpdate            GeneralizedTime,
    nextUpdate            GeneralizedTime,
    ipAddrBlocks SEQUENCE (SIZE(1..MAX)) OF ROAIPAddressFamily 
}

ASID ::= INTEGER

ROAIPAddressFamily ::= SEQUENCE {
    addressFamily OCTET STRING (SIZE (2..3)),
    addresses SEQUENCE (SIZE (1..MAX)) OF ROAIPAddress 
}

ROAIPAddress ::= SEQUENCE {
    address IPAddress,
    maxLength INTEGER OPTIONAL
}

IPAddress ::= BIT STRING
\end{lstlisting}

\section{Protobuf defintions}
\subsection{Improved Notification}\label{app:protorrdp}
Protobuf version of Notification.xml. 

\begin{lstlisting}[language=Protobuf]
syntax = "proto3";

message Notification {
  uint64 serial = 1;
  String session = 2;
  SnapshotReference snapshot = 3;
  repeated DeltaReference deltas = 4;
}

message SnapshotReference {
  String uri = 1;
  String hash = 2;
}

message DeltaReference {
  String uri = 1;
  String hash = 2;
  uint64 serial = 3;
}
\end{lstlisting}

Protobuf version of Snapshot.xml. 
\begin{lstlisting}[language=Protobuf]
syntax = "proto3";

message Snapshot {
  uint64 serial = 1;
  String session = 2;
  repeated CA cas = 43;
}

message CA {
  String repo = 1;
  repeated SnapshotEntry entries = 2;
}

message SnapshotEntry {
  String name = 1;
  bytes content = 2;
}
\end{lstlisting}

Protobuf version of Delta.xml. 
\begin{lstlisting}[language=Protobuf]
syntax = "proto3";

message Delta {
  uint64 serial = 1;
  String session = 2;
  repeated DeltaCA cas = 43;
}

message DeltaCA {
  String repo = 1;
  repeated DeltaEntry modified = 2;
  repeated DeltaEntry withdrawn = 3;
}

message DeltaEntry {
  String name = 1;
  optional bytes hash = 2;
  optional bytes content = 3;
}
\end{lstlisting}

\subsection{Improved ROA Proto}\label{app:protoroa}
Protobuf version of ROA. 
\begin{lstlisting}[language=Protobuf]
syntax = "proto3";

message ROA {
  uint64 asn = 1;
  repeated IpAndFam ip_and_fam = 2;
  Meta meta = 3;
}

message IpAndFam {
  uint32 fam = 1;
  repeated IpEntry ips = 2;
}

message IpEntry {
  bytes ip = 1;
  optional uint32 ml = 2;
}

message Meta {
  string oid = 1;
  uint64 serial = 2;
  Timestamp not_before = 3;
  Timestamp not_after = 4;
  optional bytes ski = 5;
}
\end{lstlisting}

\subsection{Improved Manifest Protobuf}\label{app:protomft}
Protobuf version of manifest. 
\begin{lstlisting}[language=Protobuf]

syntax = "proto3";


message Manifest {
  ManifestContent manifest_content = 1;
  Meta meta = 2;
  Signature signature = 3;
}
message Signature {
  string algorithm = 1;
  optional bytes parameters = 2;
  bytes signature = 3;
}

message ManifestContent {
  ManifestHashes hashes = 1;
  repeated RevokedCert revoced_certs = 2;
}

message ManifestHashes {
  string hash_algorithm = 1;
  repeated ManifestHash hash_list = 2;
}

message ManifestHash {
  string file_name = 1;
  bytes hash = 2;
}

message RevokedCert {
  uint64 serial = 1;
  Timestamp revocation_time = 2;
}

\end{lstlisting}

\subsection{RFC List}\label{app:rfclist}
The following RFCs require updates or are impact by iRPKI design. Our list excludes informational RFCs and best practices. 

\begin{itemize}
    \item RFC6480
    \item RFC6268
    \item RFC6481
    \item RFC6482
    \item RFC6486
    \item RFC6487
    \item RFC6488
    \item RFC6493
    \item RFC6818 (Update RFC)
    \item RFC5280 (Updated)
    \item RFC8360 (Not deployed)
    \item RFC8897
    \item RFC9286 (Update RFC)
    \item RFC9582 (Update RFC)
    \item RFC9589
    \item RFC8182
\end{itemize}
\end{appendix}

\end{document}